\documentclass[useAMS,usenatbib]{mn2e}
\usepackage{amscd}
\usepackage{amssymb}
\usepackage{graphicx}
\usepackage{array}
\usepackage[latin1]{inputenc}

\title[Cosmological evolution of the FRII source population]{Cosmological evolution of the FRII source population}
\author[Y. Wang and C. R. Kaiser]{Yang Wang$^{1}$\thanks{E-mail: wangyang@astro.soton.ac.uk} and Christian R. Kaiser$^{1}$\\
\\
$^{1}$School of Physics and Astronomy, University of Southampton, United Kingdom}
\begin{document}

\date{}

\pagerange{\pageref{firstpage}--\pageref{lastpage}} \pubyear{2008}

\maketitle

\label{firstpage}

\begin{abstract}
By combining a model for the evolution of the radio luminosity of an individual source with the radio luminosity function, we perform a multi-dimensional Monte-Carlo simulation to investigate the cosmological evolution of the Fanaroff-Riley Class II radio galaxy population by generating large artificial samples. The properties of FRII sources
are required to evolve with redshift in the artificial samples to fit the observations. Either the maximum jet age or the maximum density of the jet environment or both evolve with redshift. We also study the distribution of FRII source properties as a function of redshift. From currently available data we can not constrain the shape of the distribution of environment density or age, but jet power is found to follow a power-law distribution with an exponent of approximately -2. This power-law slope does not change with redshift out to $z=0.6$. We also find the distribution of the pressure in the lobes of FRII sources to evolve with redshift up to $z\sim1.2$.
\end{abstract}

\begin{keywords}
galaxies: jets - galaxies: evolution - galaxies: intergalactic medium - galaxies: luminosity function, mass function - method: statistical

\end{keywords}

\section{Introduction}

The formation and evolution of massive galaxies are important components in understanding how the universe evolved. It is not clear why AGN formation takes place at the centres of some galaxies, but it is believed that this occurs as part of the formation of all massive galaxies and jet activity is often associated with AGN. Some of the most distant and powerful active galaxies have been observed in the radio band and radio astronomy has been closely connected with cosmology ever since.

Radio galaxies with extended lobes are believed to influence a significant fraction of the matter-filled universe \citep{b53, b65,b66}. From observations in X-rays, we know there are significant interactions between radio sources and the surrounding gas on scales of several tens to hundreds of kpc from the central AGN \citep[e.g.][]{b16,b54}. A denser gas environment may imply a more massive galaxy \citep{b46}, while more massive galaxies also contain more massive black holes at their centres which can generate more powerful jets \citep[e.g.][]{b35}. Therefore we can expect that there is a connection between the jet properties and their environments. The aim of our work is to study the behaviour of the properties of the radio jets as a function of their environments and of cosmological redshift.

\citet{b22} split extragalactic radio sources into two classes depending on their large-scale morphologies: FRI and FRII. FRI objects have bright cores and edge-darkened lobes while FRII objects are edge-brightened and contain hotspots. FRIIs are more luminous than FRIs with a transition luminosity around $P_{178\rm{MHz}} \sim 10^{25}$\,W\,Hz$^{-1}$\,sr$^{-1}$. \citet{b48} found that the transition in radio luminosity extends into the optical and FRIIs are preferentially associated with more optically luminous galaxies. However, the break luminosity between the FR classes increases with increasing optical luminosity of the host galaxies. The physics leading to the FRI-FRII transition is not well understood, but may depend on the gas density in the environment \citep[e.g.][hereafter KB07]{b1,b34}. In our work, we only concentrate on the FRII-type objects, because we have good analytical models for the evolution of individual objects.

Various published samples of radio sources observed at low radio frequencies, such as 3CRR \citep{b38}, 6CE \citep{b52} and 7CRS \citep{b42}, that are complete down to their flux-limits. These samples indicate that the comoving density of radio galaxies was higher during the quasar era around redshift $z=2$ as compared to the present epoch \citep[e.g.][]{b29,b62,b27}. The
star and galaxy formation rate was also considerably higher in the quasar era. The radio luminosity function (RLF) has been developed as a tool to describe the densities of radio sources. \citet[hereafter W01]{b62} generated an RLF at 151\,MHz which shows a peak at around $z=2$. Optical and hard X-ray observations of powerful AGN also reveal a similar trend \citep[e.g.][]{b59}.

Many analytical models for extragalactic classical double radio sources have been published which characterize radio sources in terms of their dynamics and radio luminosity evolution. In this paper we use the model of \citet[hereafter KA97]{b31}, which showed that the lobes inflated by the jets emerging from the AGN grow in a self-similar fashion. The radio lobe luminosity evolution is given by \citet[hereafter KDA]{b32}. The radio synchrotron emission of the lobes is due to relativistic electrons spiralling in the magnetic field of the lobe. The model of KDA self-consistently takes into account the energy losses of these electrons due to the adiabatic expansion of the lobes, synchrotron radiation and inverse Compton scattering of cosmic microwave background photons off the electrons. \citet[hereafter BRW]{b15} and \citet[hereafter MK]{b41} essentially follow the KDA prescription, but differ in the way the relativistic particles are injected from the jet into the lobe and in the treatment of loss terms and particle
transport. Thus the radio luminosity evolution from these three models show significant differences. In our work, we will generate artificial samples by using these three models and compare the results with each other.

One of the most important tools to study the evolution of radio sources is the $P$-$D$ diagram introduced by \citet{b56}, which uses the two fundamental properties of radio sources: their radio luminosity, $P$, and linear size, $D$. \citet{b5} pointed out that it is analogous to the Hertzsprung-Russell diagram for stars. The source distribution in the $P$-$D$ plane is a result of the intrinsic evolution of individual sources and the cosmological
evolution of the source population as a whole. Since we may assume that source lifetimes are considerably shorter than the age of the universe, the $P$-$D$ diagram has been used to place some constraints on the evolution of individual sources \citep{b5,b45} and to look for consistency between data and models (KDA; BRW; MK). In our work, we will investigate the evolution of radio sources across cosmological epochs, therefore we take into account redshift at the same time and extend to a 3-dimensional $P$-$D$-$z$ data cube.

The purpose of this work is to use the evolutionary model of individual radio sources together with the RLF to generate artificial samples with large numbers of sources. From these artificial samples we can find the best fitting parameters describing the radio sources and their environments, how the jet properties are distributed and how they evolve over cosmological time scales. Our approach differs from that of \citet[hereafter KA99]{b33} who assume a 'birth function' to describe the probability of the radio source progenitors becoming active and turning into radio sources. Their intrinsic luminosity evolution is determined by the properties of their jets and the environments the
progenitors are located in at some cosmological epoch. We extend this work by directly using the RLF from W01 as a description of the birth function.

\citet{b14} also investigate the trends of radio galaxy properties with redshift. The main difference between
their work and ours is that they use a model for the radio hotspots leading to steeper tracks in the P-D diagram. They also assume a 'birth function' to investigate the distribution of the whole population. \citet[hereafter BW06]{b6} and \citet[hereafter BW07]{b7} test the same three evolutionary models for FRII sources we use here. They show that none of them fit the observational data, but again they only take into account the 'birth function' instead of the RLF. The use of a birth function in KA99, BRW, BW06 and BW07 is designed to provide a good fit of the respective model to the high-luminosity end of the RLF. Our approach uses the RLF directly without the need for a 'good guess' at the birth function. We compare our results with those of BW06 and \citet{b14}.

In section 2 we describe the RLF generated by W01. In section 3 we describe the observational samples we use in our work. We then describe the $P$-$D$-$z$ data cube we use to compare the samples in section 4. Key features of the KA97 and KDA models are discussed in section 5 and we summarize our Monte-Carlo method in section 6. In section 7 we introduce the 3-dimensional KS test which we use in the sample comparison. We show our simulation results in section 8 and compare the results from different radio emission models in section 9. A discussion of the simulation results is performed in section 10 and finally we give our main conclusions in section 11. Throughout the paper we are using a
cosmological model with $H_{0}=71$\,km\,s$^{-1}$\,Mpc$^{-1}$, $\Omega_{\rm{M}}=0.3$, $\Omega_{\Lambda}=0.7$.

\section{Radio luminosity function at 151\,MHz}

To construct our artificial samples we need to know the relative number of objects with a given radio luminosity at a given redshift. This information is provided by the radio luminosity function (RLF) defined as the number of radio sources per unit co-moving volume and per unit logarithm to base ten of luminosity at a given redshift, $\rho(P,z)$. Several determinations of $\rho(P,z)$ at various observing frequencies are available in the literature. The analytical model we use to connect source parameters with observable source properties models the radio emission of the lobes without the hotspots at the end of the jets. The emission from the hotspots is most important at high frequencies and may even dominate the total emission. To minimize the effect of the hotspot emission we use the RLF at 151\,MHz compiled by W01 on the basis of 356 sources from the 3CRR \citep{b38}, 6CE \citep{b52} and 7CRS \citep{b42} flux-limited samples.

W01 model the RLF as the sum of two distinct populations allowing independent redshift evolution. The low-luminosity population, $\rho_{\rm l}$, contains a mixture of FRI-type sources and the lowest luminosity FRII-type objects. The high luminosity population, $\rho_{\rm h}$, contains only FRII-type sources. The total RLF is then $\rho(P,z) = \rho_{\rm l} + \rho_{\rm h}$.

The low-luminosity population is modelled as a Schechter function,
\begin{equation}
\rho_{l}=f_{\rm l}(z)\rho_{\rm l0}\left(\frac{P}{P_{\rm
1\bigstar}}\right)^{-\alpha_{\rm 1}}\exp\left(\frac{-P}{P_{\rm
1\bigstar}}\right),
\end{equation}
where $\rho_{\rm l0}$ is a normalization term. At luminosities $P$ below the break luminosity $P_{\rm l\bigstar}$ the RLF approximates a power-law with slope $-\alpha_{\rm l}$. The low-luminosity population decreases exponentially above the break. The normalization of $\rho_{\rm l}$ evolves with redshift through
\begin{equation}
 f_{\rm l}(z)=(1+z)^{k_{\rm 1}}
\end{equation}
up to a maximum redshift $z_{\rm l0}$ beyond which $f_{\rm l}$ remains constant. Here we use model C of W01 and so we adopt $\log \rho_{\rm l0}=-7.523$, $\alpha_{\rm l}=0.586$, $\log P_{\rm l\bigstar}=26.48$, $k_{\rm l}=3.48$ and $z_{\rm l0}=0.710$.

The high-luminosity population is parameterized in a similar way as
\begin{equation}
 \rho_{\rm h}=f_{\rm h}(z)\rho_{\rm h0}\left(\frac{P}{P_{\rm h\bigstar}}\right)^{-\alpha_{\rm h}}\exp\left(\frac{-P_{\rm h\bigstar}}{P}\right),
\end{equation}
where the exponential cut-off is now located below the break luminosity $P_{\rm h\bigstar}$ and the power-law with slope $-\alpha_{\rm h}$ extends above the break. The number of sources in the high-luminosity population is modelled as rising up to $z=z_{\rm h0}$ and then decreasing at higher redshifts as
\begin{eqnarray}
f_{\rm h} & = & \exp \left[ - \frac{1}{2} \left( \frac{z-z_{\rm h0}}{z_{\rm h1}} \right)^2 \right] \ {\rm for} \ z < z_{\rm h0} \nonumber\\
f_{\rm h} & = & \exp \left[ - \frac{1}{2} \left( \frac{z-z_{\rm
h0}}{z_{\rm h2}} \right)^2 \right] \ {\rm for} \ z \ge z_{\rm h0}.
\end{eqnarray}
The relevant constants are $\log \rho_{\rm h0}=-6.757$, $\alpha_{\rm h}=2.42$, $\log P_{\rm h\bigstar}=27.39$, $z_{\rm h0}=2.03$, $z_{\rm h1}=0.568$ and $z_{\rm h2}=0.956$.

\begin{figure}
\includegraphics[width=0.5\textwidth]{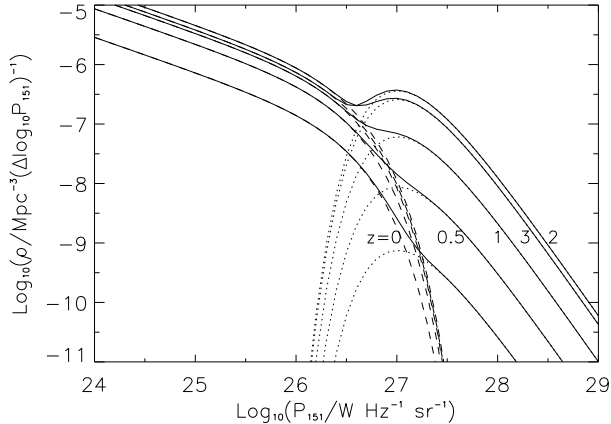}
\caption{The adopted radio luminosity function corresponding to model C of
W01 for $\Omega_{M}=0.3$, $\Omega_{\Lambda}=0.7$ and $\Omega_{k}=0$.
Dashed and dotted lines show the low-luminosity and high-luminosity
population respectively. The solid lines show the sum of both
components.} \label{rlf}
\end{figure}

In this paper we only model sources of type FRII. W01 do not distinguish between the FR classes in their determination of the RLF. While $\rho_{\rm h}$ contains only FRII-type objects, the exact composition of $\rho _{\rm l}$ in terms of FR class is not known. Here, we simply assume that 40\% of the sources contributing to the low luminosity part of the RLF are of type FRII. We find below that this assumption allows for a good fit of the properties of our artificial samples to those of the observed samples. However, the fraction of FRII-type sources in  $\rho_{\rm l}$ may be a function of redshift and/or luminosity. In fact, in Section 8.4.2 below we show that the observed sample with the lowest flux limit we use in this paper, the 7CRS sample, requires the FRII fraction to increase with increasing redshift. The advantage of using the RLF directly rather than a birth function is that we can include the weaker FRIIs in the low-luminosity population while the birth functions used by BRW, BW06 and BW07 only consider the high-luminosity population which exclusively contains FRIIs.

W01 compute the RLF for a cosmological model with $H_0=50$\,km\,s$^{-1}$\,Mpc$^{-1}$, $\Omega_{\rm M}=0$, $\Omega_{\Lambda}=0$ and $\Omega_{\rm k}=1$. We adopt the cosmological parameters consistent with the WMAP results,
$H_0=71$\,km\,s$^{-1}$\,Mpc$^{-1}$, $\Omega_{\rm M}=0.3$, $\Omega_{\Lambda}=0.7$ and $\Omega_{\rm k}=0$. Hence we
need to convert the RLF, $\rho$, to the correct cosmological model using the relation \citep{b50}:
\begin{equation}
\rho_{1}(P_{1},z)\frac{{\rm d}V_{1}}{{\rm
d}z}=\rho_{2}(P_{2},z)\frac{{\rm d}V_{2}}{{\rm d}z},
\end{equation}
where $P$ is the luminosity derived for a measured flux and redshift $z$ in a specific cosmological model, while $V$ is the comoving volume. The indices refer to two different cosmological models.

The comoving distance in a given cosmological model is \citep[e.g.][]{b28}
\begin{equation}
 D_{M}(z)  =  \frac{c}{H_{0}}\int_{0}^{z}\frac{{\rm d}z'}{\sqrt{\Omega_{\rm M}(1+z')^{3}+\Omega_{\rm k}(1+z')^2+\Omega_{\Lambda}}}.
\end{equation}
The measured flux of a source at redshift $z$ corresponds to different luminosities in different cosmological models and they are related by
\begin{equation}
 P_1 D_{M,1}^{-2} = P_2 D_{M,2}^{-2}.
\end{equation}
The comoving volume is
\begin{equation}
 {\rm d} V = 4 \pi D_{M}^2 \, {\rm d} D_{M}.
\end{equation}
We use the above relations to translate the RLF of W01 into our adopted cosmological model. We show the adopted RLF in
Figure \ref{rlf}.

\section{Complete samples of radio sources}

\begin{figure}
\includegraphics[width=0.5\textwidth]{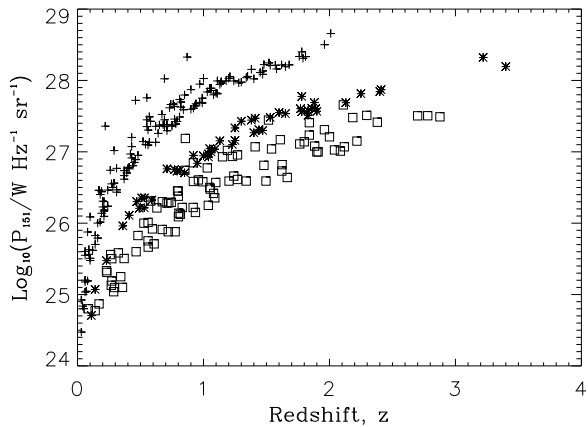}
\caption{The radio luminosity-redshift plane for the 3CRR, 6CE and
7CRS samples. The different symbols identify sources from different
samples: 3CRR (pluses), 6CE (asterisks) and 7CRS (squares).}\label{sample}
\end{figure}

Flux-limited samples contain all radio sources within a well-defined sky area with radio fluxes above a specified limit at the observing frequency. The term complete sample refers to a flux-limited sample of radio sources in which the cosmological redshifts of the host galaxies and the linear size of the lobes of all sample members are also measured. Finally, since the model used here only describes sources with FRII-type morphologies, we also need to know the classification of each sample member in terms of FR class. W01 used three complete samples, 3CRR, 6CE and 7CRS, for the determination of the RLF. We use the same samples here. However, in addition to the measured redshifts and radio luminosities of all sample members necessary for the construction of the RLF, we will also use their measured linear sizes. The luminosity-redshift distribution of the FRII-type objects contained in the three samples is shown in Figure \ref{sample}. Note the range in radio luminosity covered by the samples at a given redshift. Below we will also use the complete BRL sample \citep{b10} to estimate the goodness of fit of our models. We do not directly use the BRL sample to constrain our models because its flux limit is defined at 408\,MHz, significantly higher than the observing frequency of the RLF we use to construct our models. Below we briefly describe the samples we use in this paper.

The Third Cambridge Revised Revised sample \citep[3CRR;][]{b38} covers the largest sky area ($4.23$\,sr) of all the
samples used here, but it is also the shallowest. Its flux-limit of $S_{178} > 10.9$\,Jy is defined at 178\,MHz rather than 151\,MHz, the observing frequency of the RLF we use. The sample contains 146 FRII-type sources. To convert the flux limit as well as the fluxes of individual objects from 178\,MHz to 151\,MHz W01 simply assume a
typical spectral index of -0.8 and we follow this prescription. The flux limit therefore translates to 12.4\,Jy at 151\,MHz. We use the redshifts, angular sizes and luminosities of sample members listed online by Willott at
www.science.uottawa.ca/$\sim$cwillott/3crr/3crr.html.

The Sixth Cambridge sample (6CE) defined by \citet{b20} is based on the 6C survey. It was re-selected and updated by \citet{b52}. The radio flux at 151\,MHz of sources in this sample falls in the range $2.0\leq S_{151}\leq 3.93$\,Jy. Note that 6CE is the only sample that has an upper as well as a lower flux limit. The area covered by the sample is $0.103$\,sr. There are 52 FRII-type sources in this sample. We use the redshifts and luminosities of sample members as published in \citet{b52}, while the angular sizes of the sources are taken from \citet{b20}.

The Seventh Cambridge Redshift Survey (7CRS) is a combination of the sub-divisions I,II and III of the original 7C survey \citep{b42}. The sample is observed at $151$\,MHz and contains every source with a flux-density $S_{151}\geq0.5$\,Jy in three regions of the sky covering a total of 0.022\,sr. The data relevant for this work for the third field of the sample can be found in Table 9 of \citet{b36}. The data for the first and second fields of the
sample is referred to, but not published, by \citet{b27} and can be obtained from www-astro.physics.ox.ac.uk/$\sim$sr/grimes.html. The classification of sources in the 7C-I and 7C-II sub-fields of
the sample into FR types has not been published.

We obtained the relevant data from the VLA archive to repeat the FR classification. The raw data was reduced using standard routines within AIPS. Within the 7C-I and 7C-II fields we find 11 sources of type FRI, 52 sources with an FRII-type morphology and 11 compact sources. The compact sources are unresolved in all of the archival data down to an angular size of 0.5\,arcsec. To prevent problems in the construction of our artificial samples, we exclude any
artificial sources with linear sizes corresponding to this smallest resolvable angular size or smaller. We list our source classification for the 7C-I and 7C-II fields in Table 1 and 2.

\begin{table*}
 \centering
 \begin{minipage}{140mm}
\caption{Data for the 7C-I sub-field. Column 1 is the source name,
column 2 is the redshift, column 3 is the linear size in kpc, column
4 is the logarithm of the radio luminosity at 151\,MHz and column 5
is the morphology classification with 1 indicating FRI, 2 indicating
FRII. c indicates an unresolved, compact object. Column 6 is the
program code of the VLA observations used for the FR
classification.}
\begin{tabular}{cccccc}
\hline \hline

Name & z & Linear size (kpc) & $\rm{log}_{10}(P_{151}/\rm{WHz}^{-1}\rm{sr}^{-1})$ & Morphology & Program code\\
\hline
5C6.5 & 1.038 & 186.4 & 26.31 & 2 & AB371 \\
5C6.8 & 1.213 & 47.4 & 26.83 & c & AR335 \\
5C6.17 & 1.05 & 396.1 & 26.56 & 2 & AB371 \\
5C6.19 & 0.799 & 68.3 & 26.51 & 2 & AR335 \\
5C6.24 & 1.073 & 11.1 & 26.68 & 2 & AB766 \\
5C6.25 & 0.706 & 198.2 & 26.08 & 1 & AL355 \\
5C6.29 & 0.72 & 92.4 & 25.94 & 2 & AL355 \\
5C6.33 & 1.496 & 124 & 26.65 & 2 & AB371 \\
5C6.34 & 2.118 & 66.4 & 27.13 & 2 & AL355 \\
5C6.39 & 1.437 & 214.5 & 26.59 & 1 & AL355 \\
5C6.43 & 0.775 & 31.6 & 26.16 & 1 & AB371 \\
5C6.62 & 1.45 & 271 & 26.93 & 1 & AL355 \\
5C6.63 & 0.465 & 370.6 & 25.66 & 2 & AR477 \\
5C6.75 & 0.775 & 112 & 25.94 & 2 & AR365\\
5C6.83 & 1.8 & 119 & 27.2 & 2 & AR365 \\
5C6.78 & 0.263 & 1459.7 & 25.62 & 2 & AB667 \\
5C6.95 & 2.877 & 113 & 27.55 & 2 & AR335 \\
5C6.160 & 1.624 & 53.5 & 26.88 & 2 & AB371 \\
5C6.201 & 0.595 & 76.9 & 26.12 & 1 & AR365 \\
5C6.214 & 0.595 & 216.4 & 25.98 & 2 & AR477 \\
5C6.217 & 1.41 & 103.1 & 27.13 & 2 & AR335 \\
5C6.233 & 0.56 & 48.3 & 26.07 & 2 & AL355 \\
5C6.237 & 1.62 & 23.8 & 27.11 & c & AR365 \\
5C6.239 & 0.805 & 616.4 & 26.2 & 2 & AB766 \\
5C6.242 & 1.9 & 42.6 & 27.06 & 2 & AR365 \\
5C6.251 & 1.665 & 50.7 & 26.7 & 2 & AR365 \\
5C6.258 & 0.752 & 2.4 & 26 & c & AR365 \\
5C6.264 & 0.831 & 40.6 & 26.28 & 2 & AL355 \\
5C6.267 & 0.357 & 23.7 & 25.16 & 2 & AB766 \\
5C6.279 & 0.473 & 183.6 & 25.55 & 1 & AR365 \\
5C6.282 & 2.195 & 8 & 27.04 & c & AR365 \\
7C0221+3417 & 0.852 & 140.8 & 26.8 & 2 & AL355 \\
5C6.286 & 1.339 & 140.3 & 26.65 & 2 & AL355 \\
5C6.288 & 2.982 & 7.3 & 27.6 & c & AR335 \\
5C6.287 & 2.296 & 103.9 & 27.57 & 2 & AR335 \\
5C6.291 & 2.91 & 4.4 & 27.57 & c & AB667 \\
5C6.292 & 1.241 & 41.4 & 26.72 & 2 & AB371 \\
\hline
\end{tabular}
\end{minipage}\label{7c1}
\end{table*}
\begin{table*}
 \centering
 \begin{minipage}{140mm}
 \caption{Same as Table 1, but for the 7C-II sub-field.}
\begin{tabular}{cccccc}
\hline \hline
Name & z & Linear size (kpc) & $\rm{log}_{10}(P_{151}/\rm{WHz}^{-1}\rm{sr}^{-1})$ & Morphology & Program code\\
\hline
5C7.7 & 0.435 & 13.4 & 25.58 & 1 & AL355 \\
5C7.8 & 0.673 & 320.3 & 26.36 & 2 & AB371 \\
5C7.9 & 0.233 & 440.8 & 25.37 & 2 & AB371 \\
5C7.10 & 2.185 & 170.3 & 27.54 & 2 & AB371 \\
5C7.15 & 2.433 & 16.4 & 27.36 & c &  AR335 \\
5C7.17 & 0.936 & 691.5 & 26.21 & 2 & AL0401 \\
5C7.23 & 1.098 & 235.2 & 26.63 & 2 & AB371 \\
5C7.25 & 0.671 & 6.3 & 25.78 & c & AB667 \\
5C7.47 & 1.7 & 1.7 & 26.79 & c & AB371 \\
5C7.57 & 1.622 & 634.7 & 26.79 & 2 & AB371 \\
5C7.70 & 2.617 & 13.7 & 27.75 & 2 & AR365 \\
5C7.78 & 1.151 & 187.6 & 26.99 & 2 & AR365 \\
5C7.79 & 0.608 & 1863.9 & 25.77 & 2 & AL355 \\
5C7.82 & 0.918 & 358.3 & 26.28 & 2 & AL355 \\
5C7.85 & 0.995 & 227.4 & 26.64 & 2 & AL0401 \\
5C7.87 & 1.764 & 94.8 & 27.17 & 2 & AR335 \\
5C7.95 & 1.203 & 486.8 & 26.65 & 2 & AL0401 \\
5C7.106 & 0.264 & 104.8 & 25.28 & 1 & AB371 \\
5C7.111 & 0.628 & 80.4 & 26.29 & 1 & AB371 \\
5C7.118 & 0.527 & 76.3 & 26.06 & 2 & AL355 \\
5C7.125 & 0.801 & 120.2 & 26.15 & 2 & AB371 \\
5C7.145 & 0.343 & 93.2 & 25.31 & 2 & AB371 \\
5C7.170 & 0.268 & 97.1 & 25.19 & 2 & AB371 \\
5C7.178 & 0.246 & 121.6 & 25.15 & 1 & AB371 \\
5C7.194 & 1.738 & 16.8 & 27.3 & 2 & AB371 \\
5C7.195 & 2.034 & 22.1 & 27.12 & 2 & AB371 \\
5C7.205 & 0.71 & 107.3 & 26.34 & 2 & AB371 \\
5C7.208 & 2 & 146.7 & 27.27 & 2 & AL0401 \\
5C7.223 & 2.087 & 42.2 & 27.07 & 2 & AL355 \\
5C7.242 & 0.992 & 389.9 & 26.22 & c & AL355 \\
5C7.245 & 1.61 & 100.2 & 27.23 & 2 & AB371 \\
5C7.269 & 2.218 & 61.6 & 27.21 & 2 & AL355 \\
5C7.271 & 2.224 & 9.6 & 27.06 & 2 & AR365 \\
5C7.400 & 1.883 & 491.9 & 27.14 & 2 & AR365 \\
5C7.403 & 2.315 & 15.8 & 26.96 & c & AR365 \\
7C0825+2446 & 0.243 & 375.8 & 24.94 & 1 & AR365 \\
7C0825+2443 & 0.086 & 122.1 & 24.86 & 2 & AB371 \\
\hline
\end{tabular}
\end{minipage}\label{7c2}
\end{table*}

\citet{b10} define a complete sample (BRL) at an observing frequency of 408\,MHz and a flux limit of 5\,Jy. For a typical spectral index of -0.8 this flux limit at 408\,MHz translates to a limit close to that of the 3CRR sample at
178\,MHz. Thus the BRL sample allows us to estimate the similarity of samples drawn from the same parent population with similar selection criteria, but in different parts of the sky. However, given the large difference in observing frequency we do not attempt to fit the BLR sample directly with our model. The BRL sample contains 178 sources, 124 of which have an FRII-type morphology.

\section{The P-D-z data cube}

We use the RLF derived from observations to constrain the number of sources in our artificial samples with a given radio luminosity at a specified redshift. Therefore, by construction, our artificial samples agree with the two-dimensional distribution of radio luminosities, $P$, and redshifts, $z$, the RLF, or $P$-$z$ diagram, of the observed complete samples. However, the complete samples also contain the measured linear sizes of the radio lobes and therefore represent a three-dimensional data cube with the axes radio luminosity, $P$, linear size, $D$, and redshift, $z$. To agree with the observational constraints, our artificial samples must provide a good fit to the observed source distribution in the full $P$-$D$-$z$ data cube.

Previous work has concentrated on two-dimensional projections of the full data cube of varying subsets of the three samples we discuss here (e.g. Subramanian \& Swarup 1990; Neeser et al. 1995; KA99; BRW; Arshakian \& Longair 2000; W01; MK; Grimes et al. 2004; BW06 and BW07). In this paper we use the full three-dimensional data cube to constrain the models to avoid the loss of information associated with the projection process.

BRW, BW06 and BW07 also considered the radio spectral index as a fourth parameter. They found that the observed distribution of the spectral index is not well fitted. They use the same models for the evolution of radio sources that we employ in this paper. The models significantly restrict the possible range of the spectral index and this effect naturally explains the difficulties encountered by BW06 and BW07. Here we do not attempt to reproduce the observed distribution of spectral indices as this would either result in the same problem or require the introduction of an additional model parameter. We will return to this point below.

\section{Models for the dynamic and emission evolution of individual sources}

The large-scale structure of FRII-type sources is inflated by powerful jets accelerated in the vicinity of the supermassive black holes at the centre of the AGN inside the host galaxies. The jets end in strong shocks which accelerate electrons to relativistic velocities and may increase the strength of the magnetic field. The magnetized plasma passing through the jet shock subsequently inflates the lobe or cocoon. The basic dynamical picture was first
proposed by \citet{b55} and \citet{b13}.

\citet{b21} and KA97 showed that if the jet is in pressure-equilibrium with its own lobe, then the expansion of the
lobe and the bow shock in front of it is self-similar. Here we use the model of KA97 to describe the dynamics of the lobes. We summarize the most important features of this model below.

The growth of the jet length is essentially determined by a balance of the ram pressures of the jet material and that of the medium surrounding the host galaxy which is pushed aside by the jet. X-ray observations of the hot gas in the vicinity of elliptical galaxies, in galaxy groups and in galaxy clusters find density distributions well fitted by $\beta$-models \citep[e.g.][]{b24}. In KA97 the density distribution outside the core radius $a_0$ is approximated by a power-law:
\begin{equation}
\rho_{x}=\rho_{0}\left(\frac{d}{a_{0}}\right)^{-\beta},
\end{equation}
where $d$ is the radial distance from the AGN at the centre of the density distribution and $\rho_0$ is the density at the core radius. The exponent $\beta$ is constrained by observations to the range $0<\beta \le 2$.

We assume that the gas in the vicinity of the AGN has a non-relativistic equation of state, $\Gamma _x = 5/3$, while the lobes only contain magnetic fields and relativistic particles, $\Gamma _c =4/3$. For a jet providing a constant power $Q_0$ for a time $t$ and inflating a lobe with an aspect ratio $R_{\rm T}$ we then find the lobe length, $D$, and the pressure inside the lobe, $p_{\rm c}$,
\begin{eqnarray}
D & = & c_{1}\left(\frac{Q_{0}}{\rho_{0}a_{0}^{\beta}}\right)^{\frac{1}{5-\beta}} t^{\frac{3}{5-\beta}}, \nonumber\\
p_{\rm c} & = & \frac{27 c_{1}^{2-\beta}}{16(5-\beta)^{2}}R_{\rm T}^{-2}(\rho_{0}a_{0}^{\beta})^{\frac{3}{5-\beta}}t^{\frac{-4-\beta}{5-\beta}}Q_{0}^{\frac{2-\beta}{5-\beta}},
\label{pressure}
\end{eqnarray}
where
\begin{equation}
c_{1}=\left(\frac{64R_{\rm T}^{4}(5-\beta)^{3}}{81\pi(8-\beta+3R_{\rm T}^{2})}\right)^{\frac{1}{5-\beta}}.
\end{equation}
Note here that the model of the source dynamics only depends on the combination $\rho_0 a_0^{\beta}$, but not on $\rho_0$ and $a_0$ separately. For convenience we therefore introduce the parameter $\Lambda = \rho_0 a_0^{\beta}$. In our model we initially assume $\beta=2$ as this considerably simplifies the model equations. We will discuss the model with $\beta=1.5$ later.

KDA97 extended the dynamics model of KA97 to include the calculation of the radio emission expected from the lobe. The model takes into account the energy losses of the relativistic electrons due to the adiabatic expansion of the lobe, synchrotron radiation and inverse Compton losses from scattering of the cosmic background radiation. These losses modify the energy distribution of the relativistic electrons which we assume to initially follow a power-law with exponent $p$ between a low and high energy cut-off represented by the Lorentz factors of the least and most energetic electrons, $\gamma_{\rm min}$ and $\gamma_{\rm max}$, respectively. To simplify the equations and the model we choose $p=2$ for all sources. We also set $\gamma_{\rm min} =1$ and $\gamma_{\rm max} =10^5$. Note that the exact value of the power-law limits does not significantly influence the model results as long as $\gamma_{\rm min} \ll
\gamma_{\rm max}$. A summary of all relevant model equations can be found in KB07.

Our choice of $p$ limits the range of spectral indices predicted by the model to the narrow range from $-0.5$ to $-1$. For young sources, synchrotron energy losses are most important, while for old objects inverse Compton losses dominate and both processes lead to a steep spectrum. For `middle-age' sources between these extremes, the relativistic electrons are not much affected by radiative energy losses and so the spectrum flattens (KB07). In
objects with observed spectral indices not conforming to this pattern a different acceleration regime may be applicable and lead to a different value of $p$. Our choice of a fixed value for $p$ implies that we cannot expect the spectral index distribution of our artificial source populations to agree with that of the observed complete samples. However, the introduction of an unknown distribution for $p$ into the model would considerably complicate
the analysis of our results and introduce another model parameter.

The combined model of KA97 and KDA in the form specified here depends on five model parameters, $z$, $R_{\rm T}$, $\Lambda$, $t$ and $Q_0$. A complete set of these parameters fully determines the linear size of the lobes, $D$, and their radio luminosity, $P$. Therefore we can take the observed distribution of sources in the $P$-$D$-$z$ space to constrain the underlying distribution of the density of their environments described as $\Lambda$, their age $t$
and their jet power $Q_0$. In the following we explain the practical implementation of this process as well as our assumption for the distribution of lobe aspect ratios, $R_{\rm T}$.

\section{Monte-Carlo simulation}

\begin{figure}
\includegraphics[width=0.5\textwidth]{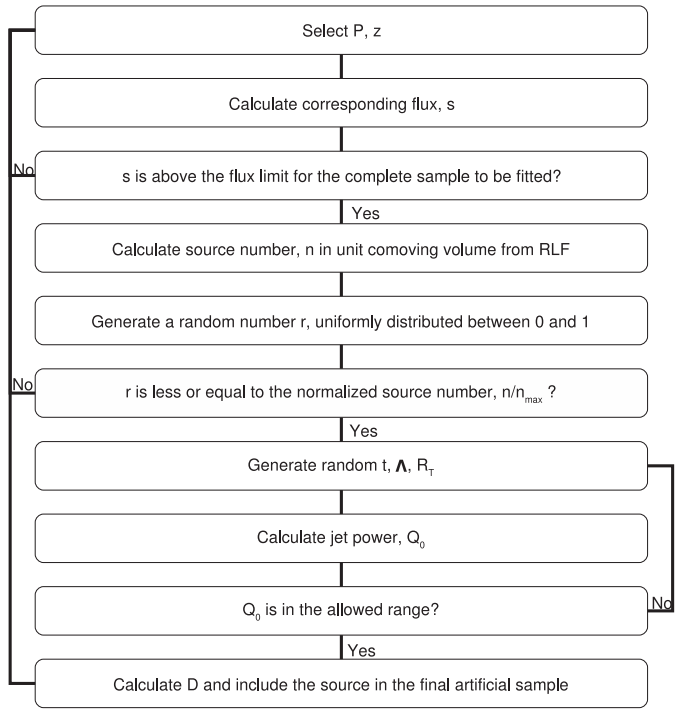}
\caption{The flow diagram of the Monte-Carlo process we use in this paper.}\label{flow}
\end{figure}

The RLF, $\rho \left( P, z \right)$, gives the comoving number density of radio sources at a cosmological redshift $z$ with a given radio luminosity $P$. The number of sources within the ranges $z$ to $z+{\rm d} z$ and $\rm{log}P$ to $\rm{log}P+{\rm d}\rm{log}P$ is given by $n=\rho \left( P, z \right) \, {\rm d} \log P \, {\rm d} V$, where ${\rm d} V$ is the comoving volume sampled between $z$ and $z+{\rm d} z$. The relevant formulae are presented in Section 2. We consider redshifts from $z=4$ to $z=0$ with a step size of ${\rm d} z =0.004$ and radio luminosities at an observing frequency of 151\,MHz from $\log P =24.5$ to $\log P = 29$ in steps of ${\rm d} \log P =0.01$. These ranges cover the entire source population in all three observed samples we consider. We only calculate $n$ for combinations of $P$ and $z$ for which the corresponding observable flux exceeds the flux limit of the observed complete sample we use to compare our artificial sample with.

After calculating $n$ at all positions defined by our ranges and step sizes, we normalize $n$ such that its maximum is equal to unity. We can now interpret $n \left(P, z\right)$ as the distribution of the probability to find a source in an artificial sample with a given combination of redshift and radio luminosity. To construct our artificial sample we then iterate through all allowed combinations of $P$ and $z$ and generate a random number, $r$, uniformly distributed between 0 and 1. If $r\le n \left(P, z\right)$ for the chosen $P$ and $z$, then a source with this radio
luminosity and redshift is included in the artificial sample. We can increase the sample size by repeating the same process a fixed number of times at all allowed combinations of $P$ and $z$.

Once a source with a given combination of radio luminosity and redshift is included in the artificial sample, we need to calculate the linear size of its lobe to determine its position in the $P$-$D$-$z$ data cube for comparison with the observations. For this, we need the aspect ratio of the lobe, $R_{\rm T}$, the age of the source, $t$, and the parameter describing the external density distribution, $\Lambda$. Fixing the values of these three parameters
uniquely determines the length of the lobes, $D$, and the jet power, $Q_0$. In other words, fixing the values of three parameters uniquely determines the values of the remaining two. For each source in the artificial sample we will choose random values for $R_{\rm T}$, $t$ and $\Lambda$ from distributions of these parameters discussed in the following in order to assign values to their linear sizes and jet power.

It is difficult to accurately determine the aspect ratio of the lobe of a radio source. The lobe should ideally be detected all the way back to the central AGN which suggests the use of a low observing frequency to minimize the effects of radiative energy losses. At the same time we need sufficient angular resolution to resolve the lobe perpendicular to the jet axis. This is best achieved at high observing frequencies. In practice a balance must
be found between these requirements and so an observationally determined distribution of $R_{\rm T}$ is not available in the literature. For simplicity we choose a uniform distribution across the range $1.3 \le R_{\rm T} \le 6$, where the limits are motivated by observed values (Leahy \& Williams, 1984). Below we discuss the minor effects on our results of changing this distribution.

If we assume that all radio sources have a common maximum lifetime $t_{\rm max}$, then the distribution of their ages is given by a uniform distribution extending from $t=0$ to $t= t_{\rm max}$. Spectral ages of some objects reach a few $10^8$\,years \citep[e.g.][]{b2}, but recently \citet{b64} found an average age of around $10^7$\,years for a sample of FRII-type objects located in low redshift galaxy groups, suggesting a maximum lifetime of a few $10^7$\,years. The maximum lifetime depends on the availability of fuel for the jet-producing AGN and it would therefore be surprising, if $t_{\rm max}$ was the same for all sources. However, the spread around the average age found by \citet{b64} is comparatively small and for simplicity we assume that $t_{\rm max}$ is indeed the same for all sources at a given redshift. We will show below that $t_{\rm max}$ defined in this way may be a function of redshift.

The parameter $\Lambda$ describing the density distribution in the environment of a source can, in principle, be determined from X-ray observations. FRII-type objects seem to be preferentially located in isolated galaxies or galaxy groups.  With $\beta=2$, we expect $\Lambda$ between $10^{17}$\,kg\,m$^{-1}$ and $10^{18}$\,kg\,m$^{-1}$ for individual galaxies \citep{b24}. For group environments \citet{b30} find values for $\Lambda$ in the range $2\times 10^{18}$\,kg\,m$^{-1}$ and $4 \times 10^{19}$\,kg\,m$^{-1}$. Note that all these determinations predict
somewhat high values for $\Lambda$, because we are using $\beta =2$ at the moment. For $\beta=1.5$ the values of $\Lambda$ are more consistent with the values given by \citet{b24} and \citet{b30}. The gas haloes of many galaxies and galaxy groups show flatter slopes implying smaller values of $\Lambda$. Again there is no determination of the complete distribution of $\Lambda$ for the environments of radio sources available in the literature. Hence we draw random values for $\Lambda$ from a uniform distribution extending from $\Lambda =0$ to $\Lambda = \Lambda_{\rm max}$, where the maximum value is to be determined from the models and may be a function of redshift (see below).

For every allowed combination of the radio luminosity and the redshift plus the randomly chosen values for $R_{\rm T}$, $t$ and $\Lambda$ we can now proceed to calculate a linear size $D$ of the lobe and jet power $Q_0$ for each source to be included in the artificial sample. The parameter distributions used here allow for a wide range of possible values for $Q_0$. Not all values for $Q_0$ are acceptable. Sources with weak jets, i.e. small jet powers, will develop turbulent, rather than laminar jets and this will result in FRI-type lobes. \citet{b51} suggest that the
transition between the FR types occurs close to $Q_0 = 10^{37}$\,W based on radio observations of lobes and optical line emission from the AGN itself. A similar lower limit for the jet power of sources with an FRII-type morphology was derived by KB07 using the model we employ here as well. An upper limit on $Q_0$ may be given by the Eddington luminosity of the most massive black holes. In our model we require that $10^{37}\,{\rm W} \le Q_0 \le 5 \times 10^{40}\,{\rm W}$. Model parameters that imply a jet power outside this range are rejected. For rejected sources we generate new sets of model parameters, $R_{\rm T}$, $t$ and $\Lambda$, in the way described above until an acceptable value for $Q_0$ is found.

Note here that the mathematical form of the distributions for the model parameters only indirectly influences the distributions of these parameters in the final artificial sample. Combinations of model parameters must lead to acceptable results in the sense that not only must the resulting source have the correct radio luminosity and redshift, its jet power must also fall within the specified range. The resulting parameter distributions will in general not follow the distributions they are originally drawn from, because many possible parameter combinations will be rejected as described above. In large enough artificial samples the choice of uniform distributions over other mathematical functions therefore only weakly influences the final distributions of the model parameters.

Once an acceptable combination of model parameters is found, we can calculate the corresponding length of the lobes, $D$. This is the physical length of the lobe and in order to compare to the observed complete samples we need to take into account that the lobes may be projected as their main axis is oriented at an angle $\theta$ to our line of sight. The observed lobe length is therefore given as $D_{\rm ob} = D \sin \theta$, where the random orientation $\theta$ is distributed as $\sin \theta \, {\rm d} \theta$ between $\theta =0$ and $\theta = \pi / 2$. Before including a source in the $P$-$D$-$z$ data cube of our artificial sample, we choose a random orientation angle and `project' the lobes into the sky plane. For simplicity, hereafter we denote the $D_{\rm ob}$ as just $D$.

The process of generating our artificial samples described above is summarized in the flow diagram in Figure \ref{flow}. Our artificial samples contain of order 6000 sources. This is a much larger number than is contained in the observed samples. However, we found that such a large number is required to arrive at reasonably smooth source distributions in the $P$-$D$-$z$ data cubes and also to minimize the influence of the initial model parameter
distributions on the final distributions.

\section{Kolmogorov-Smirnov test}

Our artificial samples define a source distribution within the $P$-$D$-$z$ data cube equivalent to the observed samples. We use the three-dimensional version of the Kolmogorov-Smirnov (KS) test to compare our model results with the observations.

The classical one-dimensional Kolmogorov-Smirnov test makes use of the probability distribution of the quantity $D_{\rm KS}$, defined as the largest absolute difference between the one-dimensional cumulative distributions of two samples, where one or both samples can be continuous or discrete. \citet{b49} extends this idea to a two-dimensional test by making use of the maximum absolute difference between two distributions, when all four possible ways to
cumulate data following the directions of the coordinate axes are considered. For the comparison of a sample with $n$ members with a continuous distribution Peacock's test requires that the cumulative distributions should be calculated in all $4n^{2}$ quadrants of the plane defined by,
\begin{eqnarray}
(x<X_{i},y<Y_{j}),\ (x<X_{i},y>Y_{j}),\ (x>X_{i},y<Y_{j}),
\nonumber\\
(x>X_{i},y>Y_{j}),
\end{eqnarray}
for all possible combinations of the indexes $i$ and $j$ from $1$ to $n$. Here, $X_i$ and $Y_i$ denote the coordinates of individual members of the discrete sample.

\citet{b23} show that it is sufficient to consider only the four quadrants defined by each individual data point in the discrete sample, i.e. $i=j$. This reduces the total number of quadrants in which the distributions are accumulated to $4n$. Furthermore, the extension of this methodology to three dimensions is straightforward, provided all eight quadrants defined by each individual data point are considered in deriving the largest difference, $D_{\rm KS}$, of the cumulative distributions.

Using the three-dimensional KS test we can now compare the fit of various artificial source distributions within the $P$-$D$-$z$ data cube arising from different models with that of the observed samples. However, the method does not assign a formal goodness-of-fit measure because the statistics of $D_{\rm KS}$ in the three-dimensional KS test is in general not known. Also, the selection criteria for the observed samples, in particular their flux limit, prevent the population of certain parts of the $P$-$D$-$z$ cube. To assess which model provides an acceptable fit to the observations, we separately construct the statistics of $D_{\rm KS}$ for each model calculation. For this we generate a large number of sources for a specific model as detailed above. We then repeatedly draw random subsamples from these model sources with a total source number equal to that of the observed comparison sample. We calculate the total difference $D_{\rm KS}$ of the cumulative distributions of the subsample and the parent model sample. In this way we build up the probability distribution of $D_{\rm KS}$ for this particular model. Based on this statistic we
assign a probability $P(D_{\rm KS})$ to the value $D_{\rm KS}$ calculated for the observed sample. $P(D_{\rm KS})$ is then the probability that the observed sample arises as a subsample drawn randomly from all radio sources in the universe, if that population is described by the model. While this technique may guide us in the selection of models, it clearly cannot identify the statistically most likely model because the distribution of $D_{\rm KS}$ will be different for each individual model.

\section{Results}

\begin{figure*}
\includegraphics[width=\textwidth]{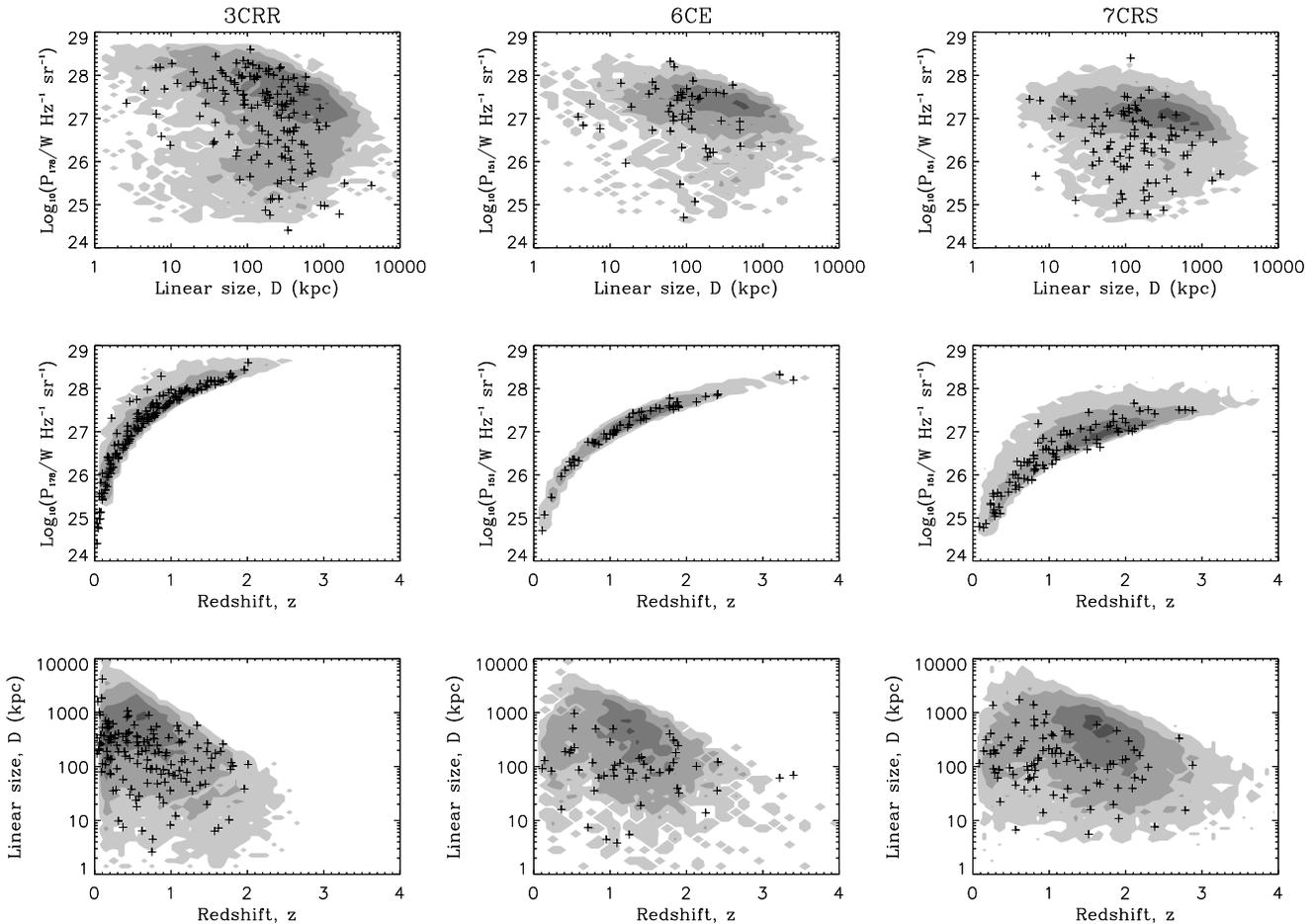}
\caption{Projections of the $P$-$D$-$z$ data cube for observed
samples with source density contours from Model A. The parameters we
use here are $t_{\rm max}=2.5\times10^{7}$\,yr and
$\Lambda_{\rm max}=1\times10^{18}$\,kg\,m$^{-1}$. Crosses indicate FRII
sources in the observed samples. Gray scales indicate the number
density of the artificial samples.}\label{modela}
\end{figure*}

\begin{figure*}
\includegraphics[width=\textwidth]{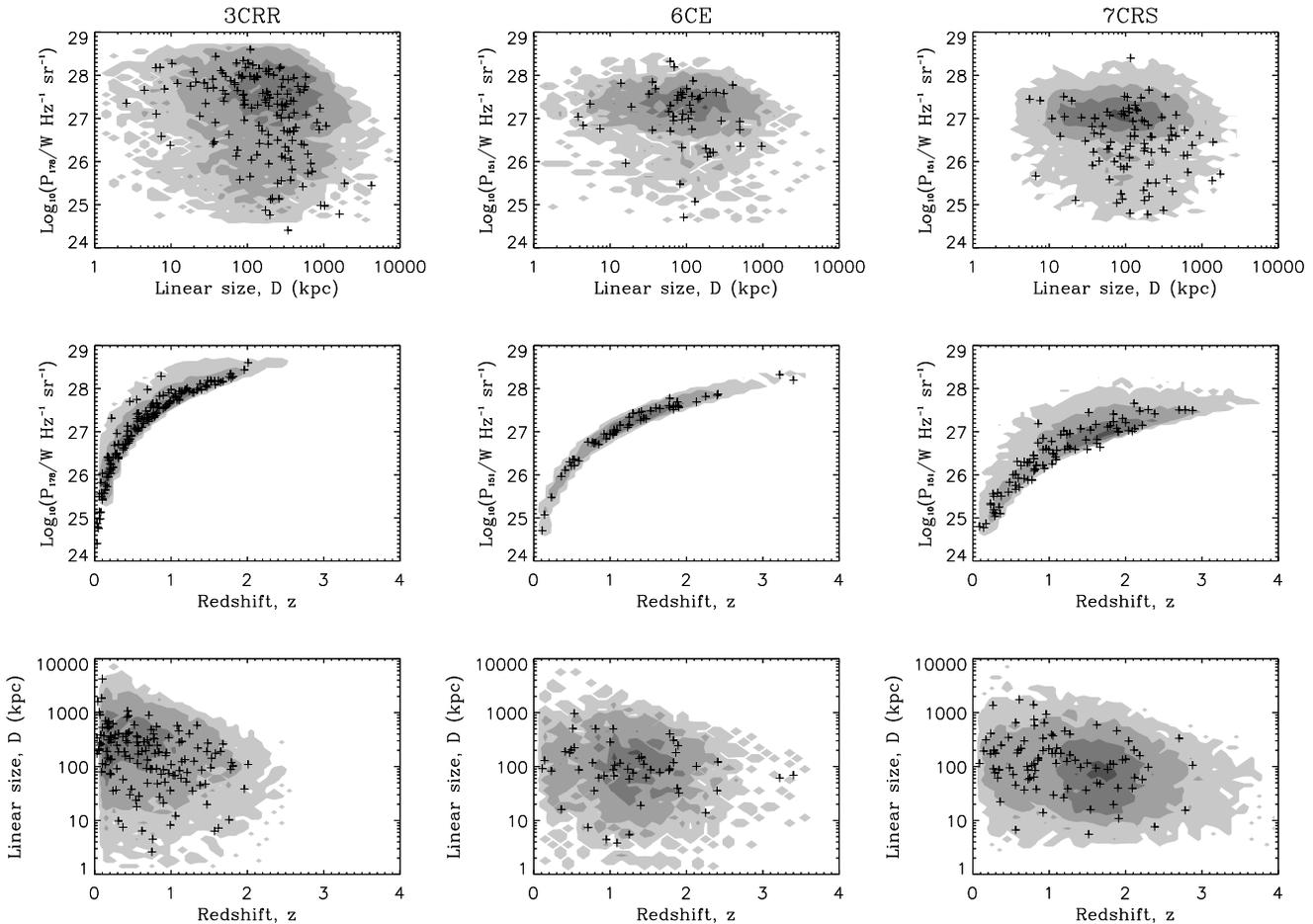}
\caption{Projections of the $P$-$D$-$z$ data cube comparing observed
samples with the source density contours resulting from Model B. The
parameters we use here are $t_{\rm max}=5\times10^{7}$\,yr,
$\Lambda_{\rm max}(0)=3.7\times10^{18}$\,kg\,m$^{-1}$ and $\psi=5.8$.
Crosses indicate FRII sources in the observed samples. Gray scales
indicate the number density of the artificial
samples.}\label{modelb}
\end{figure*}

\begin{figure*}
\includegraphics[width=\textwidth]{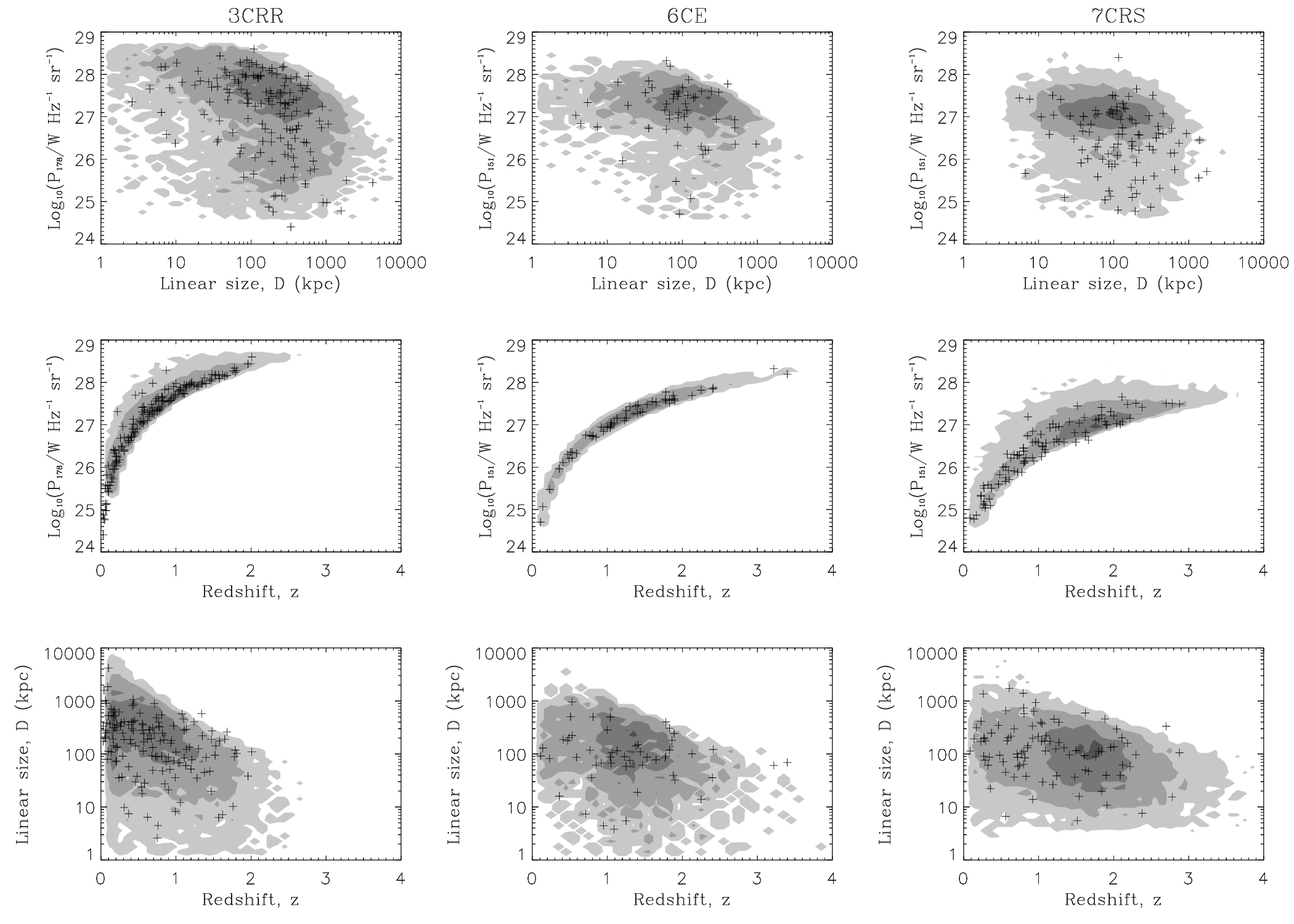}
\caption{Projections of the $P$-$D$-$z$ cube for the observed
samples with source density contours from Model C. The parameters we
use here are $t_{\rm max}(0) =2.7\times10^{7}$\,yr, $\Lambda_{\rm
max}=6.4\times10^{17}$\,kg\,m$^{-1}$ and $\phi=2.4$. Crosses
indicate FRII sources in the observed samples. Gray scales indicate
the number density of the artificial samples.}\label{modelc}
\end{figure*}

We now have all the ingredients to generate our artificial samples and compare them to the observed samples. In doing so our general approach is to start with the simplest models and only modify these as necessary to achieve a better agreement with the observations. It is not feasible to show the full three-dimensional source distributions used in the comparisons. In order to present our results, we therefore plot two-dimensional projections of the $P$-$D$-$z$ cube. Sources in observed samples are presented as individual crosses while the large numbers of sources in the artificial samples are shown as density contours in these plots. The contours enclose areas of 1\%, 10\%, 40\% and 80\% of the maximum density in each plot.

Note that in the following we always show the radio luminosity density of sources in the 3CRR sample and associated artificial samples at 178\,MHz rather than 151\,MHz. While we use the RLF at 151\,MHz in our models to calculate relative source numbers as detailed above, we calculate the luminosity for our artificial sources at 178\,MHz and use the appropriate flux limit for this frequency when comparing with the 3CRR sample.

\subsection{Model A}

The simplest model we can build within the framework described above has uniform distributions with fixed upper limits for the source age and the density parameter. We assume a fixed $t_{\rm max}=2.5 \times 10^7$\, years and investigate a range of possible $\Lambda _{\rm max}$. None of the $\Lambda _{\rm max}$ tried leads to a probability
significantly different from zero as measured by the KS test. Varying $t_{\rm max}$ instead of $\Lambda_{\rm max}$ leads to the same result.

As an example, in Figure \ref{modela} we show a comparison of the resulting artificial samples with $\Lambda _{\rm max} = 1 \times10^{18}$\,kg\,m$^{-1}$ with the observed samples. Clearly high luminosity sources located at high redshifts are too large in our artificial samples compared to the observed sources. The sources in the artificial samples show a trend of increasing size with increasing luminosity until the trend is reversed at the highest
luminosities. The radio luminosity of a source in the KDA model is mainly governed by the jet power, $Q_{0}$. However, a larger value of $Q_{0}$ also implies faster growth of the lobes and, all other parameters being equal on average, it is not surprising that we find a trend of radio luminosity with size in our artificial sample.

The reversal of this trend at the highest luminosities has been noted by many authors \citep[e.g.][]{b47,b45}
for observed samples. In the artificial samples it is caused by the flux limits of the samples in combination with the decreasing luminosity of older, and therefore larger, sources. Clearly the evolutionary model used here predicts that this effect alone is not sufficient to explain the observed trend reversal and some other effects must contribute to reconcile the model with the observations (see also KA99). We will investigate such effects in the next two sections.

Modified models with a steeper luminosity evolution of individual sources do not need to invoke such additional effects (BRW and MK). However, they may be less consistent with the overall source distribution (BW06). We
shall apply these models in the same way to the data as the combined models of KA97 and KDA in section 9.

\subsection{Model B}

In this and the following sections we investigate which additional factors may lead to a better fit of the predicted distribution of sources in the $P$-$D$-$z$ cube with the observational data. We have shown above that in our modeling framework we need an additional effect to explain the apparent shortening of the lobes of sources with the highest radio luminosities located at the highest redshifts. In this section we introduce a redshift dependence for the upper limit of the density parameter $\Lambda$. In the following section we do the same for the upper limit of the source age distribution.

We fix the maximum source age to $t_{\rm max} = 5 \times10^7$\,years and introduce a variable upper limit for the
distribution of $\Lambda$ such that $\Lambda_{\rm max} (z) = \Lambda_{\rm max} (0) \left( 1 + z \right)^{\psi}$. The best agreement between artificial and observed samples is achieved for 3CRR with $\Lambda_{\rm max}(0)=3.7\pm0.2 \times 10^{18}$\,kg\,m$^{-1}$ and $\psi=5.9\pm0.2$, giving a probability that the observed sample is drawn from a population described by the model of $P (D_{\rm KS}) = 48\%$. The values for 6CE are $\Lambda _{\rm max} (0) =
3.8^{+0.2}_{-0.3}\times 10^{18}$\,kg\,m$^{-1}$, $\psi=5.8\pm0.2$ and $P(D_{\rm KS}) = 56\%$. The limits on the model parameters give their value where $P(D_{\rm KS})$ halves compared to its maximum. The fit of the model to the 7CRS sample is much worse and the values for $P(D_{\rm KS})$ are very low. This is related to problems with the used RLF for this sample. We discuss this issue in Section 8.4.2. In any case, the maximum of $P(D_{\rm KS})$ for 7CRS occurs close to $\Lambda_{\rm max} (0) =3.7 \times 10^{18}$\,kg\,m$^{-1}$ and $\psi=6$, consistent with the results for the other two samples. A model with $\Lambda_{\rm max} (0) = 3.7 \times 10^{18}$\,kg\,m$^{-1}$ and $\psi =5.8$ provides an acceptable fit to both 3CRR ($P(D_{\rm KS})=40\%$) and 6CE ($P(D_{\rm KS})=41\%$). The comparison of this model and the observed samples is shown in Figure \ref{modelb}.

The much better fit of Model B compared to Model A is explained by the considerably higher average density of the source environments at high redshifts. The density parameter $\Lambda$ has only a moderate influence on the radio luminosity of a model source compared to the jet power. However, a high value of $\Lambda$ efficiently reduces the expansion speed of the lobes. Hence the strong evolution of $\Lambda_{\rm max}$ with $z$ in Model B ensures
that sources at high redshift remain smaller for longer than their low redshift counterparts. This allows a better fit of the high luminosity / high redshift part of the source population.

\begin{figure*}
\includegraphics[width=0.8\textwidth]{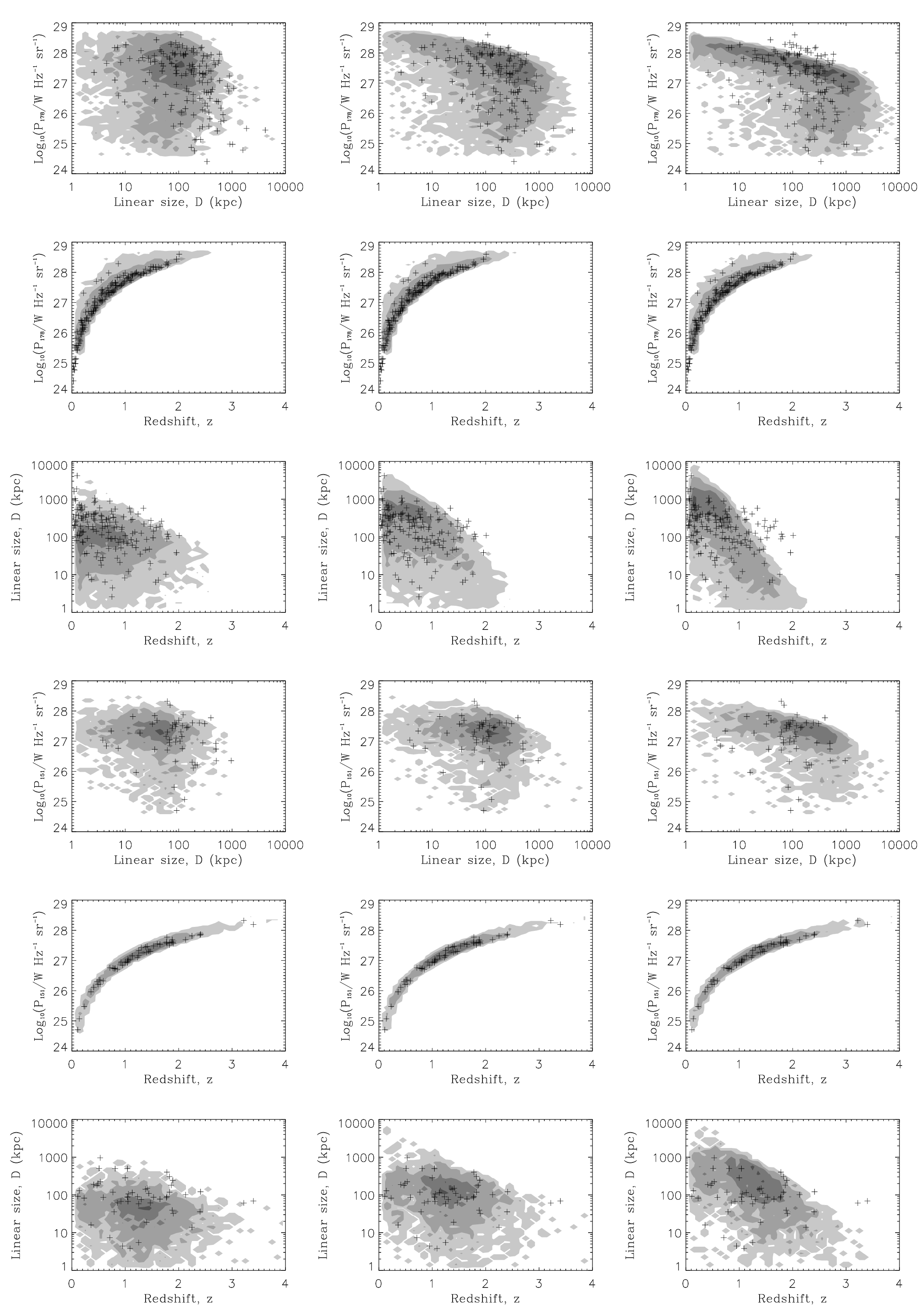}
\caption{Projections of the $P$-$D$-$z$ cube of observed and
artificial samples with fixed $R_{T}$. The parameters we use here
are those of the best fitting Model C. Here we only concentrate on
3CRR and 6CE samples as the fit to 7CRS is poor. The upper nine
panes are artificial samples corresponding to 3CRR with constant
$R_{\rm T}=1.3, R_{\rm T}=3.4$ and $R_{\rm T}=6.0$ from left to
right. The lower nine are artificial samples corresponding to 6CE
with $R_{\rm T}=1.3, R_{\rm T}=2.8$ and $R_{\rm T}=6.0$ from left to
right.}\label{rtmodel}
\end{figure*}

\begin{figure*}
\includegraphics[width=\textwidth]{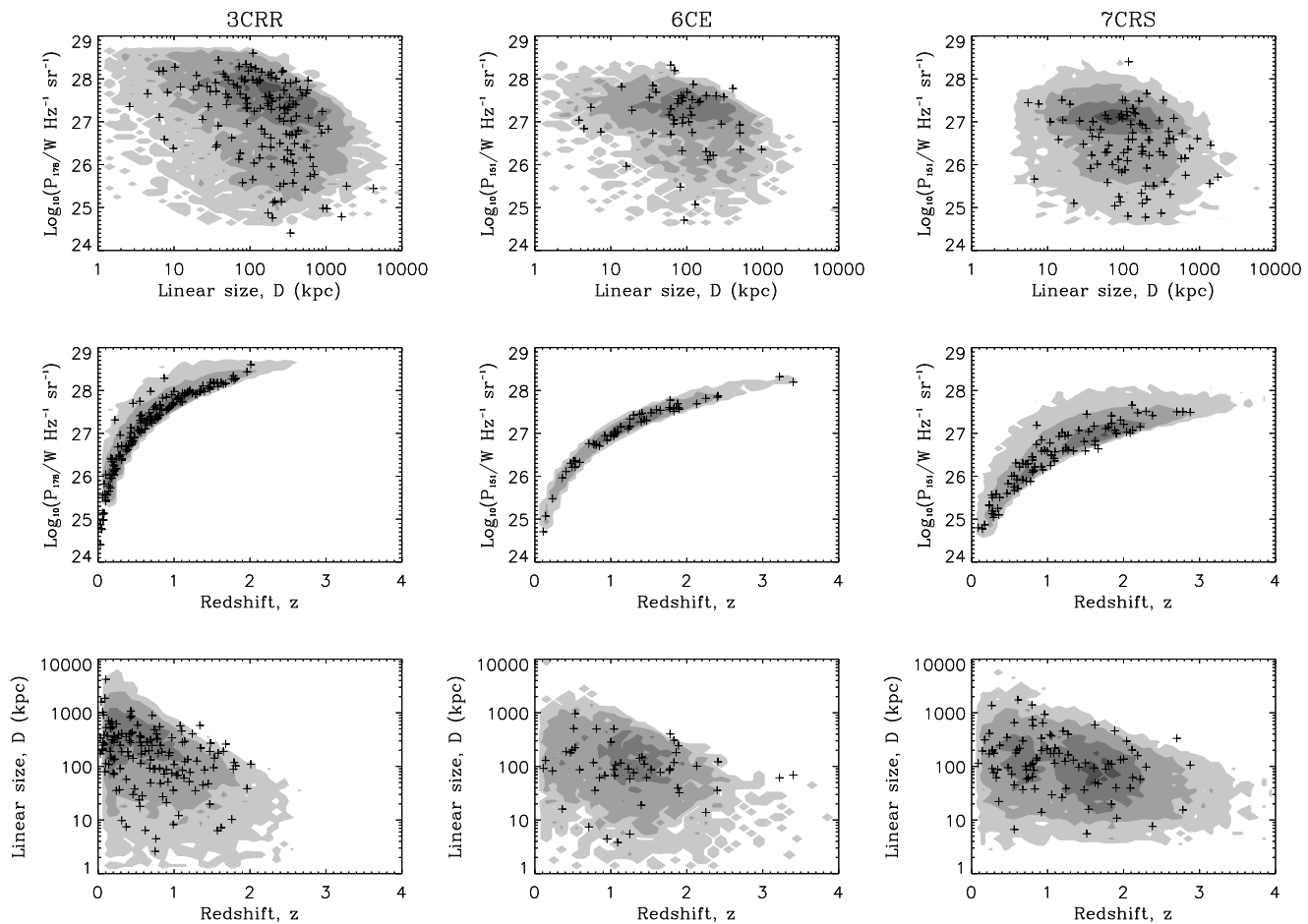}
\caption{Projections of the $P$-$D$-$z$ cube of the observed and
artificial samples with a modified low-luminosity population of the
RLF. The artificial sample is generated by Model C with the
best-fitting parameters.}\label{rlf1}
\end{figure*}

\begin{figure*}
\includegraphics[width=\textwidth]{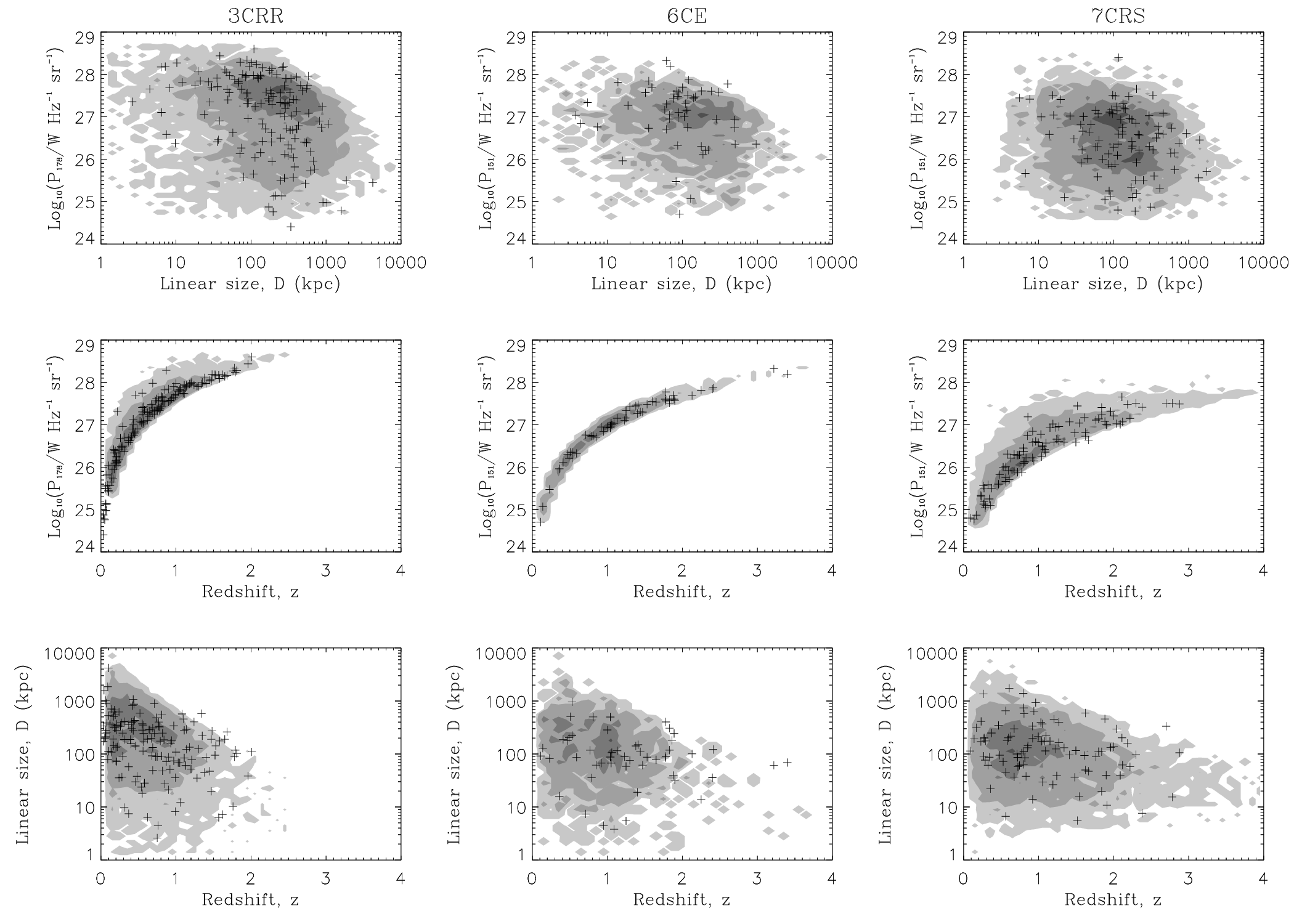}
\caption{Projections of the $P$-$D$-$z$ cube of the observed and
artificial samples with both the low and high luminosity populations
of the RLF modified. The artificial sample is generated by Model C
with the best fitting parameters.}\label{rlf2}
\end{figure*}

\subsection{Model C}

Instead of invoking denser average source environments at higher redshifts to reduce the average size of the highest luminosity sources, we can also reduce the average source lifetime for increasing redshift. For this we fix $\Lambda _{\rm max} (z)$ at $6.4 \times 10^{17}$\,kg\,m$^{-1}$ for all $z$ and introduce a variable maximum age as $t_{\rm max} (z) = t_{\rm max} (0) \left( 1 + z \right)^{-\phi}$. The best agreement is found for $t_{\rm max}(0) = 2.7^{+0.3}_{-0.2}\times10^{7}$\,yr and $\phi=2.4\pm0.2$ for 3CRR with $P(D_{\rm KS})=76\%$. The 6CE sample gives an almost identical result with $t_{\rm max}(0)=2.7^{+0.1}_{-0.3}\times10^{7}$\,yr and $\phi=2.4\pm0.1$ where $P(D_{\rm KS}) = 59\%$. Again, the comparison with 7CRS does not produce good fits for the reasons we will discuss in Section 8.4.2, but the maximum of the probability occurs close to the parameter values for the two other samples. The maximum source age at low redshifts is reassuringly close to the value of the average source lifetime of $1.2 \times 10^7$\,yr recently derived by \citet{b64} for an independent sample. Figure \ref{modelc} compares the model results with the observed data.

We assume a uniform distribution of $\Lambda$ in Model C. The real distribution of $\Lambda$ is unknown and it does not have to be uniform. Instead of an uniform distribution, we also introduce a Gaussian distribution with its peak at
$3.2\times10^{17}$\,kg\,m$^{-1}$ and a width of $\sigma=1\times10^{17}$\,kg\,m$^{-1}$. The distribution is truncated so that there are no values of $\Lambda$ below 0 or above $6.4\times10^{17}$\,kg\,m$^{-1}$. This distribution covers the same range of $\Lambda$, but is of course different from an uniform distribution. The agreement with the same $t_{\rm max}(0)$ and $\phi$ are $P(D_{\rm KS})=77\%$ for 3CRR and $P(D_{\rm KS})=52\%$ for 6CE. For the extreme assumption of a fixed value of $\Lambda=3.2\times10^{17}$\,kg\,m$^{-1}$ for all sources, we can still find a comparable agreement with $P(D_{\rm KS})=73\%$ for 3CRR and $P(D_{\rm KS})=55\%$ for 6CE. Clearly, we can not rule out these distributions compared to an uniform distribution. We will discuss this point in detail in section 10.

Both models B and C can provide adequate fits to the observed samples 3CRR and 6CE. Given the results detailed above, we cannot formally decide whether it is a reduced maximum lifetime or a denser environment that limits the lobe sizes of sources with the greatest radio luminosities at the highest redshifts accessible to the observed samples. Of course, a combination of the two effects is also not ruled out. However, the result shows that with currently
available samples it is not necessary to introduce yet more complicated models to explain the distribution of sources in the $P$-$D$-$z$ data cube. We will concentrate on Model C in the discussion of parameter dependencies, alternative models and the further implications of our results. This does not indicate that we prefer Model C over Model B, but is simply done to avoid confusion.

\subsection{Potential problems}

\subsubsection{The power-law exponent, $\beta$}

$\beta$ is the power-law exponent which indicates how fast the environment density decreases away from the center AGN. In previous sections, we use $\beta=2$, but most studies of the environments of low redshift radio galaxies in X-ray band imply that $\beta$ could be closer to 1.5 than 2. Here we apply $\beta=1.5$ in the KDA model to check how much it will affect our results. 

With a smaller value of $\beta$, the density of the environment decreases more slowly and the speed of growth of the lobe is slower. In this case, it is not surprising to have a higher maximum jet age. The best fitting parameters we find are $t_{\rm max}(0)=1.1\times10^{8}$\,yr, $\phi=2.6$ for fixed $\Lambda_{\rm max}=3.5\times10^{8}$\,kg\,m$^{-1.5}$. The godness of fit decreases slightly to $P(D_{\rm KS})=72\%$ for 3CRR and $P(D_{\rm KS})=41\%$ for 6CE. The agreement is still good while the value of $t_{\rm max}(0)$ is still consistent with the range discussed in Section 6. Therefore, the value of $\beta$ does not significantly change our simulation results and conclusions. We will still concentrate on the models with $\beta=2$ in the following sections.

\subsubsection{The axial ratio, $R_{T}$}

In our construction of the artificial samples we have used a fixed uniform distribution for the aspect ratio of the lobes $R_{\rm T}$. Since $R_{\rm T}$ affects both the size and the radio luminosity of the lobes, it is reasonable to ask how much our results depend on our assumptions for $R_{\rm T}$.

If we replace the uniform distribution of $R_{\rm T}$ over the range 1.3 to 6 we used so far with a constant $R_{\rm T}$ for all model sources, then the distribution of lobe sizes within our artificial samples becomes slightly narrower. The average size also shifts somewhat depending on the value assumed for $R_{\rm T}$. This effect is illustrated in Figure \ref{rtmodel}. There is no noticeable restriction of the luminosity range. For Model C with the best fitting parameters, but a constant value for $R_{\rm T}$, we find that the probability $P(D_{\rm KS})$ decreases to 34\% for 3CRR ($R_{\rm T} =3.4$) and remains constant at 59\% for 6CE ($R_{\rm T}=2.8$). We conclude that changing the range or distribution of $R_{\rm T}$ will only have a minor effect on our results. This is particularly encouraging because the alternative evolutionary models for individual sources we introduce below assume fixed values for $R_{\rm T}$.

\subsubsection{The RLF}

The adoption of an RLF allows us to include the low-luminosity FRII sources in the artificial sample, and is a more direct method to constrain the source distribution than the introduction of a birth function. However, in general a RLF requires more parameters than a simple 'birth function'. The changes in these parameters may also change the simulation results. Here we use the generalized luminosity function(GLF) constructed by \citet{b27} for comparision with the RLF of W01 used so far.

\citet{b27} consider both radio luminosity and optical luminosity of the AGN and introduce a parameter $\alpha$ encoding the $L_{151}-L_{\rm OIII}$ correlation and a parameter $\beta'$ encoding scatter about this correlation. The GLF based on $\alpha$ and $\beta'$ can generate a smoother RLF than the RLF of W01. However the total source counts predicted by the GLF do not change very much compared to the RLF of W01 as \citet{b27} show in their Figure 11. If we substitute the GLF into Model C with the same best fitting parameters and a modified FRII fraction of 50\% in the low-$\alpha$ population, we can still find a reasonable agreement with $P(D_{\rm KS})=34\%$ for 3CRR and $P(D_{\rm KS})=26\%$ for 6CE. The use of an RLF introduces many parameters, and the changes in these parameters do affect the fitting result. However within a reasonable range, we can still get good agreement between artificial and observed samples with a modified RLF. Meanwhile, the conclusion that the FRII properties evolve with redshift do not change with the choice of RLF.

The FRII fraction in the low-luminosity population is another parameter introduced in the RLF that can affect the fitting result. The value of $40\%$ comes from the simulation. If we apply a smaller fraction, for example $20\%$, we will find too few low-luminosity sources. The agreement for 3CRR and 6CE drops below $1\%$. Therefore, although some fainter surveys at low frequency such as the Bologna surveys indicate the FRII fraction at the low luminosities could be smaller,  we still need to choose $40\%$ as the appropriate value for our work for fitting the 3CRR, 6CE and 7CRS samples.

\subsubsection{The 7CRS sample}

In all our models discussed above we noted that our approach cannot provide an adequate fit to the source distribution of the 7CRS sample. The main problem appears to be the different relative number of sources at low redshifts and high redshifts in the various samples. This may be due to a changing mix of sources with different
FR morphology below the break in the RLF. So far we have assumed that a constant fraction of 40\% of the sources contributing to the low luminosity part of the RLF of W01 are of type FRII, irrespective of redshift. A closer look at the three observed samples reveals that this assumption may be too simplistic.

Within a range from $10^{24}$\,W\,Hz$^{-1}$\,sr$^{-1}$ to $10^{26}$\,W\,Hz$^{-1}$\,sr$^{-1}$ at 151\,MHz the 3CRR sample contains 20 sources in a redshift range $0.005 \leq z \leq 0.16$ of which 12 show an FRII-type morphology. In the same luminosity range, 7CRS contains a total of 33 sources spanning a redshift range from $z=0.086$ to $z=0.775$ of which 26 are of type FRII. These numbers suggest an increased fraction of FRII-type sources within the low luminosity population at higher redshifts. The 6CE sample does not fit into this trend with 3 FRII-type sources out of a total of 7 within the luminosity range and at redshifts intermediate between 3CRR and 7CRS. However, the number of sources in this sample is very small.

Based on these numbers we test whether a redshift dependence of the FR mix helps to improve the model fit to the 7CRS data. We set the FRII fraction within the low luminosity population of the RLF of W01 equal to $0.3 +z$ up to $z=0.7$ and keep it at unity at higher redshifts. With the modification of the RLF the agreement of Model C with the 7CRS data increases, but not significantly. The value of $P(D_{\rm KS})$ increases for 3CRR to 84\%, while it decreases for
6CE to 47\%. A comparison of the model with the observational data is shown in Figure \ref{rlf1}.

Changing the FR mix in the low luminosity part of the RLF does not have the desired effect of improving the fit with the 7CRS data. However, the presently available data do not allow us to decide whether a change with redshift in the composition of the radio source population is taking place or not.

From Figure \ref{rlf1} we see that there are too many sources in our artificial sample around $z=2$ and $P_{151} = 10^{27}$\,W\,Hz$^{-1}$\,sr$^{-1}$ compared to the 7CRS sample. In fact, this problem, that the RLF of W01 overpredicts the number of sources with this radio luminosity at this redshift compared to 7CRS, is discussed by W01 (see their Figures 7 and 8). The 7CRS data is consistent with no further evolution of the high luminosity part of the RLF beyond redshift $z=1$. Implementing such a constant RLF at redshifts beyond $z=1$ in our model dramatically improves the agreement between our Model C and the 7CRS data to $P(D_{\rm KS})=17\%$. However, this RLF reduces the agreement between Model C and the 3CRR sample to $P(D_{\rm KS})=0.1\%$ while $P(D_{\rm KS})$ for the 6CE sample drops below
0.1\%. Therefore it is unlikely that this modification of the RLF is necessary. The 7CRS sample covers a very small sky area compared to the 3CRR and 6CE samples. This may imply that the members of this sample are not fully representative of the entire source population. Comparisons of the artificial samples arising from this modification
of the RLF with the observations are shown in Figure \ref{rlf2}.

\subsection{Comparing the 3CRR sample with an equivalent sample}

\begin{figure}
\includegraphics[width=0.5\textwidth]{e_3}
\caption{3CRR and BRL objects in the P-D-z diagram at 178\,MHz. Cross
symbols refer to 3CRR while squares refer to BRL. We convert the BRL
sources from 408\,MHz to 178\,MHz by using a common spectral
indexes of 0.8}\label{e_3}
\end{figure}

\begin{figure}
\includegraphics[width=0.5\textwidth]{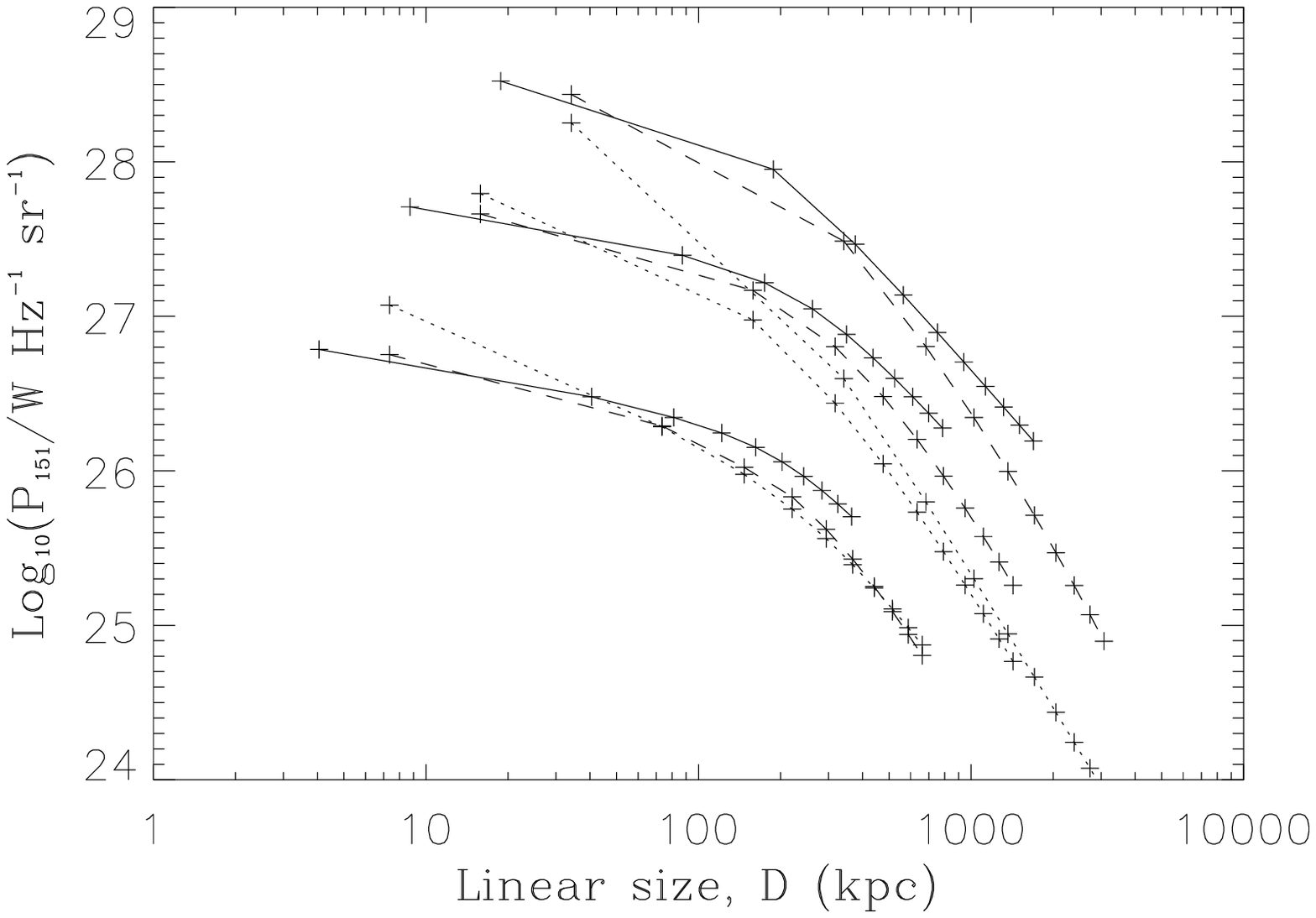}
\caption{Evolutionary track for three sources with different models.
The upper curves are for $Q_{0}=1\times10^{40}$\,W and $z=1.5$, the
centre curves are for $Q_{0}=1\times10^{39}$\,W and $z=0.5$, and the
lower curves are for $Q_{0}=1\times10^{38}$\,W and $z=0.2$. Each of
the solid, dashed and dotted curves refer to the tracks given by
KDA, MK and BRW models respectively. The pluses on the curves are
time markers denoting source lifetimes of
1,10,20,...90\,Myr.}\label{track}
\end{figure}

The BRL sample, as discussed in section 3, is roughly equivalent to 3CRR with a similar flux limit and source number. Thus the BRL sample and the 3CRR sample can be considered as two samples drawn from the same universe with similar selection criteria, but from different areas of the sky. The comparison between these two samples indicates the agreement between two samples drawn from the same parent population and can therefore give us some idea of how well
our models agree with the observations.

The BRL sample is observed at 408\,MHz, so we use a constant spectral index of $-0.8$ to convert the radio luminosity of BRL sources to 178\,MHz. We ignore the sources whose luminosity is below the flux limit given by 3CRR. Comparing the 3CRR sample and the BRL sample in the $P$-$D$-$z$ data cube, we get $D_{\rm{KS}}=0.2662$ from the 3-D KS test. If we choose an artificial BRL sample with the same source number as the real BRL sample from the best-fitting artificial 3CRR sample and compare with the real 3CRR sample, $D_{\rm{KS}}$ is mostly between $0.1$ to $0.2$. As the smaller values of $D_{\rm{KS}}$ indicate a better fit, this shows that our best-fitting artificial sample agrees with the observations to a degree similar to the agreement between similar observed samples drawn from the same source population. Figure \ref{e_3} shows the 3CRR sample and the BRL sample in the $P$-$D$-$z$ data cube.

\section{Comparison between models of radio lobe evolution}

\begin{figure*}
\includegraphics[width=\textwidth]{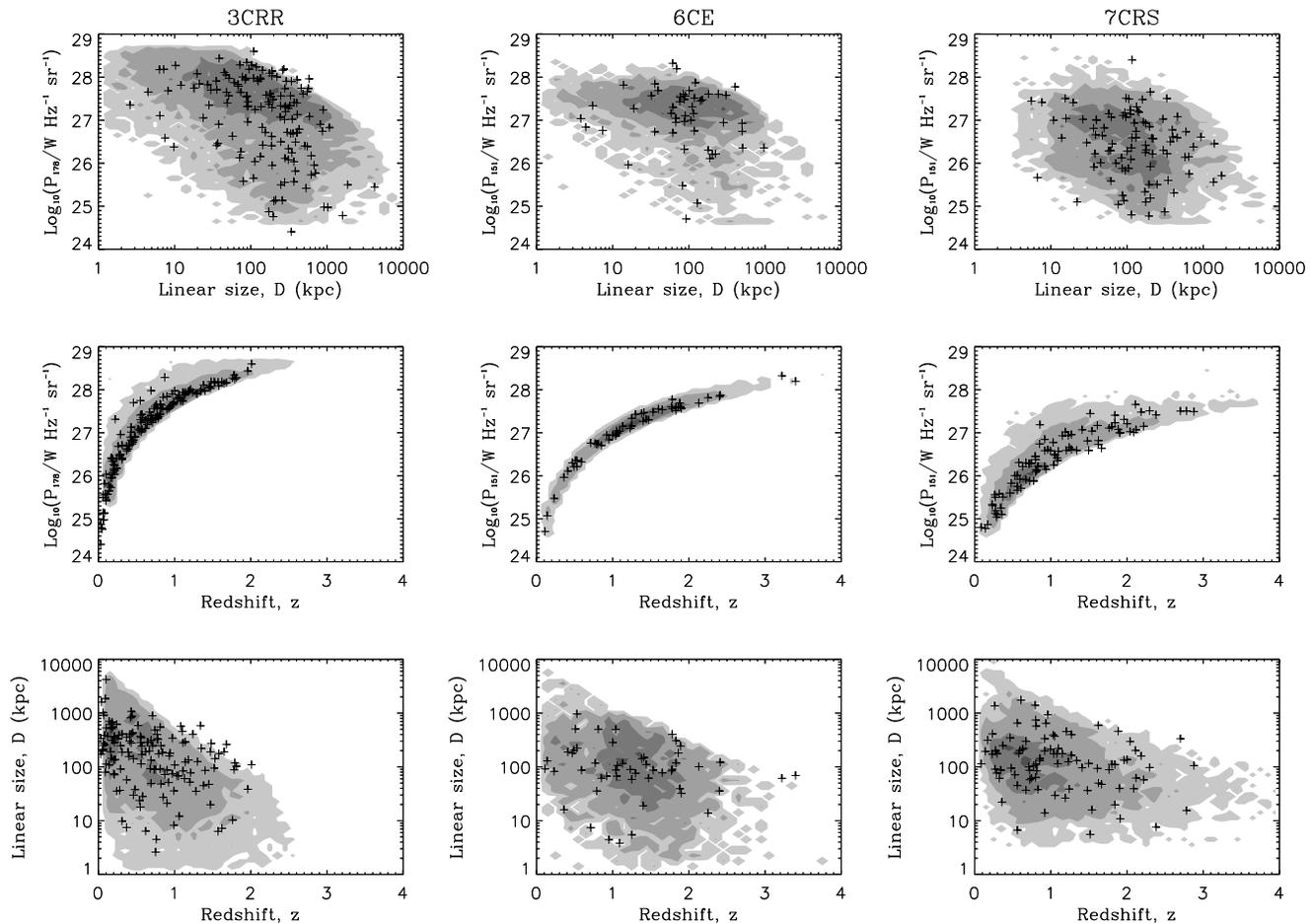}
\caption{Projections of the $P$-$D$-$z$ cube of the observed samples
with source density contours generated by the BRW model in the
framework of Model C. The parameters we use here are $t_{\rm
max}(0)=5.0\times10^{7}$\,yr, $\Lambda_{\rm
max}=1.6\times10^{18}$\,kg\,m$^{-1}$ and $\phi=2.5$. Crosses
indicate FRII sources in the observed samples. Gray scales indicate
the number density of the artificial samples.}\label{brw}
\end{figure*}

\begin{figure*}
\includegraphics[width=\textwidth]{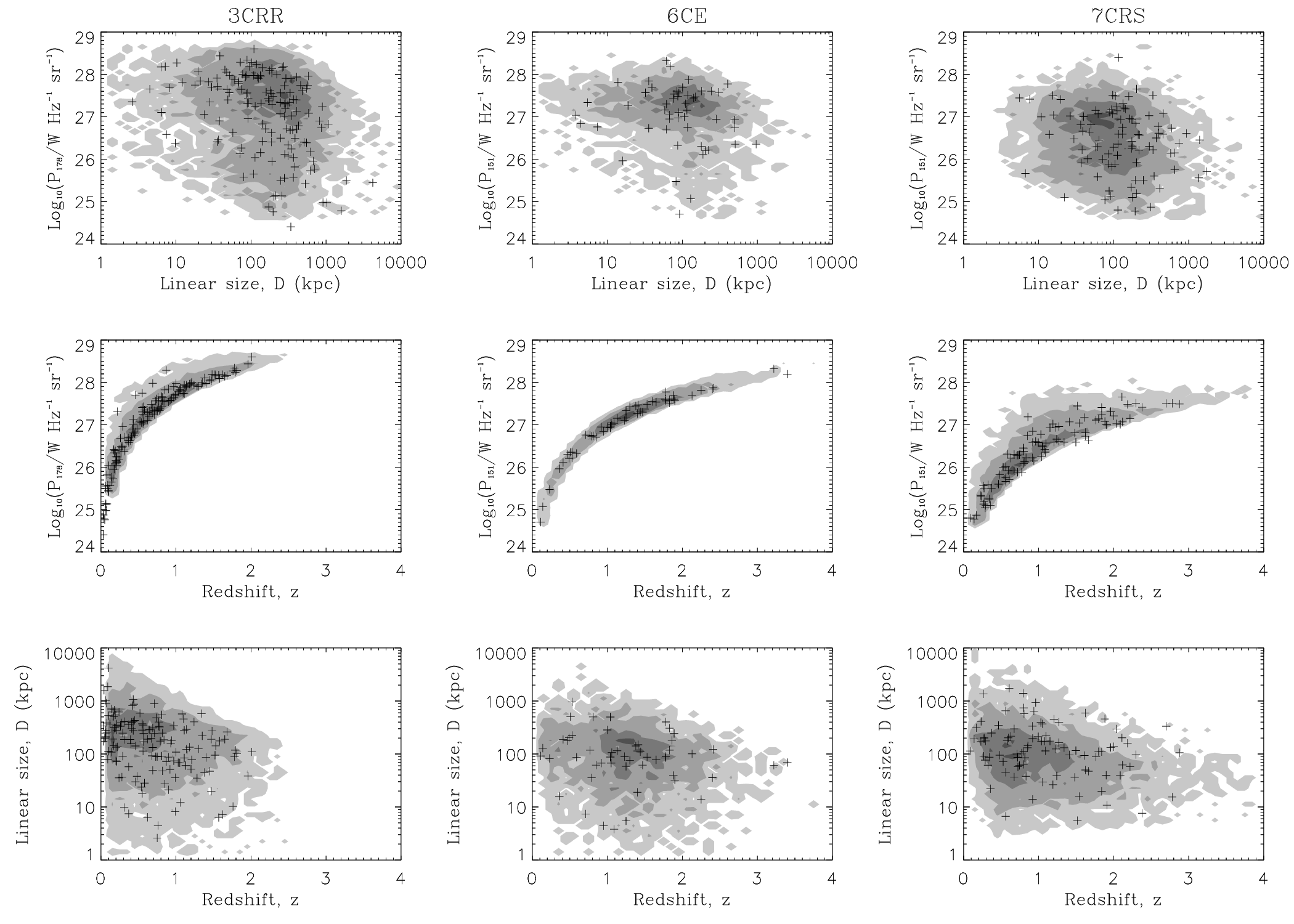}
\caption{Projections of the $P$-$D$-$z$ cube of the observed samples
with source density contours generated by the MK model within the
framework of Model C. The parameters we use here are $t_{\rm
max}(0)=4.0\times10^{7}$\,yr, $\Lambda_{\rm
max}=1.6\times10^{18}$\,kg\,m$^{-1}$ and $\phi=2.5$. Crosses
indicate FRII sources in the observed samples. Gray scales indicate
the number density of the artificial samples.}\label{mk}
\end{figure*}

Before discussing our results in more detail, we now assess how much they depend on our particular choice for the model of the evolution of individual sources. The model describing the dynamics and expansion of the radio lobe, KA97, essentially relies on the condition of ram pressure balance between the jet material and the receding ambient gas in front of it. This condition was first introduced by \citet{b55} and has formed the basis for virtually all subsequent models of the dynamical evolution of radio sources with an FRII-type morphology \citep[e.g.][]{b9,b21,b44,b18}. While the details of the derivation of the lobe dynamics differ, the basic principle is the same. As we do not consider in detail the evolution of individual objects, we therefore do not use another dynamical source model.

Based on the dynamical model of KA97, two alternative descriptions for the emission properties of the lobes have been formulated by BRW and MK. The main difference between these models and KDA is the treatment of the radiating electrons as they propagate from the ends of the jets, or hotspots, into the lobes. The KDA model assumes that the energy distribution of the electrons is described by a power-law and that they only suffer adiabatic losses during the transition from the hotspot into the lobe. BRW considers the effect of synchrotron energy losses within the hotspot by introducing a doubly broken power-law for the energy distribution. MK describe in detail the diffusion of the electrons from the hotspot into the lobe taking into account radiative energy losses and possible re-acceleration.
Figure \ref{track} shows a comparison of the luminosity evolution of a single radio source as described by the three models. The luminosity evolution is steeper for the two alternative models with the BRW model the steepest. In the following we will describe both models in more detail and then substitute them for the KDA model in our formalism. BW06 and BW07 use the same models in their comparisons.

When comparing the results arising from the use of the three different evolutionary models, we employ Model C with $\Lambda_{\rm max} = 1.6 \times 10^{18}$\,kg\,m$^{-1}$, consistent with the original formulation of BRW and MK. However, to allow a direct comparison with our previous results we set $\beta =2$, $\gamma_{\rm min}=1$ and $\gamma_{\rm max}=10^5$. Also note that the two models in their original form require a fixed value for $R_{\rm T}$. As shown in Section 8.4.1, this should not lead to major changes of the results. Finally, we also want to compare the model predictions with the 7CRS data. In order to do so we use the modified RLF as described in Section 8.4.2. For the 3CRR and 6CE samples we continue to use the unmodified RLF.

\subsection{The BRW model}

The BRW model takes the lobe dynamics from the KA97 model. This sets the pressure in the lobe, $p_{\rm c}$, according to equation (\ref{pressure}). As in KDA, the pressure in the hotspot in front of the end of the jet is higher than $p_{\rm c}$ by a constant factor. With the assumption of energy equipartition in the hotspot, the hotspot pressure fixes the strength of the magnetic field and the normalization of the energy distribution of the electrons at a given
point in time. BRW assume further that the hotspot has a fixed physical size and that it takes electrons between 1\,yr and $10^5$\,yr to escape from the hotspot into the lobe, independent of any other model parameter. These timescales give rise to two breaks in the electron energy spectrum. Their positions are calculated using the standard equipartition formulae. Below the low energy break the electron energy spectrum is modeled as a power-law with an exponent of $p=2$. Above the high energy break, $p=3$ is assumed while in between the breaks any value $2\le p \le 3$ can be adopted. In our implementation of the BRW model we use $p=2.5$ in the intermediate region. After the electrons escape into the lobe, they preserve their double power-law energy distribution, but the breaks in the power-law evolve with time.

Figure \ref{brw} shows the result of replacing the KDA model in our method with the BRW model. The best agreement of Model C is achieved with $t_{\rm max} (0) = 5.0 \times 10^7$\,yr and $\phi = 2.5$. The artificial sample contains too few large sources at high redshift compared to 3CRR. The agreement is very low at $P(D_{\rm KS}) \sim 0.1\%$. The steep luminosity evolution predicted by the BRW model leads to older and therefore large sources dropping quickly below the flux limit of the observed samples, in particular at high redshifts. In a sense the BRW `over-solves' the problem of the lack of large objects at high redshift. This was also noted by MK and BW06.

The results for the 6CE sample and the 7CRS sample, with the modified RLF, are better, since these samples contain fewer large sources at high redshifts. However, the ratio of small to large sources is still too large at high redshifts. The agreement for 6CE is $P(D_{\rm KS})=19\%$ and for 7CRS $P(D_{\rm KS}) \sim 0.1\%$, in both cases lower than for the KDA model. We conclude that the steep luminosity evolution predicted by the BRW model provides a worse fit to the observational data than the flatter evolution arising from the KDA model.

\subsection{The MK Model}

The MK model is similar to the BRW model as it also assumes that the jets end in hotspots of a constant physical size which also contain a magnetic field of constant strength. The radiating electrons are accelerated to an initial power-law energy distribution within the hotspot. In contrast to the BRW model, MK exactly follow the subsequent evolution of the energy distribution under the influence of adiabatic and radiative energy losses as the electrons diffuse through the hotspot into the lobe. MK adopt a diffusive transport model in which the mean square distance traveled by individual electrons on their way from the hotspot into the lobe is proportional to $t^{\alpha}$, where $t$ is the time since their acceleration. A value of $\alpha =1$ corresponds to the diffusion regime and we adopt this here, since the sub-diffusion regime, $\alpha < 1$, makes the model comparable to the BRW model and the
supra-diffusion regime, $\alpha > 1$, approximates the KDA model.

The MK model takes into account adiabatic and radiative losses of the electrons during the diffusion process. However, MK find that the adiabatic losses lead to luminosity evolution of their sources in disagreement with observations. To avoid this problem we adopt their model B, with all corresponding parameter settings, which
neglects adiabatic losses. We also follow MK in setting the ratio of the diffusive transport time and the cooling time of an electron, their parameter $\tau$, to $2\times 10^{-3}$. After the electrons have diffused into the lobe, the MK model describes the further evolution of the electron energy distribution in the same way as the KDA model. An example for the luminosity evolution using the MK model in the way described here is shown in Figure \ref{track}.

Replacing the KDA model with the MK model in our method we find the best fitting Model C with the parameters $t_{\rm max}(0)=4.0\pm0.2\times10^{7}$\,yr and $\phi=2.6\pm0.1$ for 3CRR at $P(D_{\rm KS} )=53\%$. For 6CE we find $P(D_{\rm KS})=58\%$ for $t_{\rm max} (0) =4.2\pm0.3\times10^{7}$\,yr and $\phi=2.5\pm0.1$. The values of the model parameters almost agree between the two samples and Figure \ref{mk} shows a comparison of the artificial and observed samples for $t_{\rm max} (0) = 4.0 \times 10^7$\,yr and $\phi = 2.5$ resulting in $P(D_{\rm KS} ) =41\%$ for 3CRR, $P(D_{\rm
KS}) =50\%$ for 6CE and $P(D_{\rm KS} ) \sim 1\%$ for 7CRS, with the modified RLF. The fitting result for 7CRS is not as good as that for the KDA model, because the evolutionary tracks of the MK model are steeper and predict more sources with small linear size at the high redshift / high luminosity end.

For the 3CRR and 6CE samples, the MK model provides a level of agreement between the artificial and observed samples similar to that of the KDA model. The somewhat higher value for $t_{\rm max}(0)$ we find for the MK model is partially due to the large value of $\Lambda _{\rm max}$ we use for this model compared to the KDA model. However, the steeper luminosity evolution of the MK model also requires somewhat longer lifetimes of the sources to accommodate the larger objects. With the currently available data we cannot decide which of the two models provides a better description. We continue to focus on the KDA model because its mathematical formulation is simpler than that of MK.

\section{Discussion}

We produce artificial samples of radio-loud AGN with an FRII-type morphology and compare their properties with those of samples of observed objects. The artificial samples are consistent with the observed ones provided that there is some cosmological evolution of the radio source population. Our models require either that the density of the source environments increases on average for increasing cosmological redshift, or that the lifetime of the jet flows decreases with increasing redshift. A combination of both effects may also be at work. To simplify the following discussion of the properties of the artificial samples, we concentrate on model C, implying the cosmological evolution of the jet lifetimes. However, we do not consider this model superior to Model B.

The steeper luminosity evolution of the radio lobes described by the model of BRW leads to a worse agreement of the model predictions with observations. The model proposed by MK provides comparable results to those of the KDA model. BW06 come to a similar conclusion. BW07 consider modified models taking into account the increase of hot-spot sizes with jet lengths. The modified BRW and MK models produce better fits which are at least as good as the KDA model, while the modified KDA model produces a worse fit. Here we do not attempt to construct a new, improved model for individual FRII-type objects and so we concentrate on the original KDA model which is considerably simpler and mathematically less complex.

\begin{figure}
\includegraphics[width=0.5\textwidth]{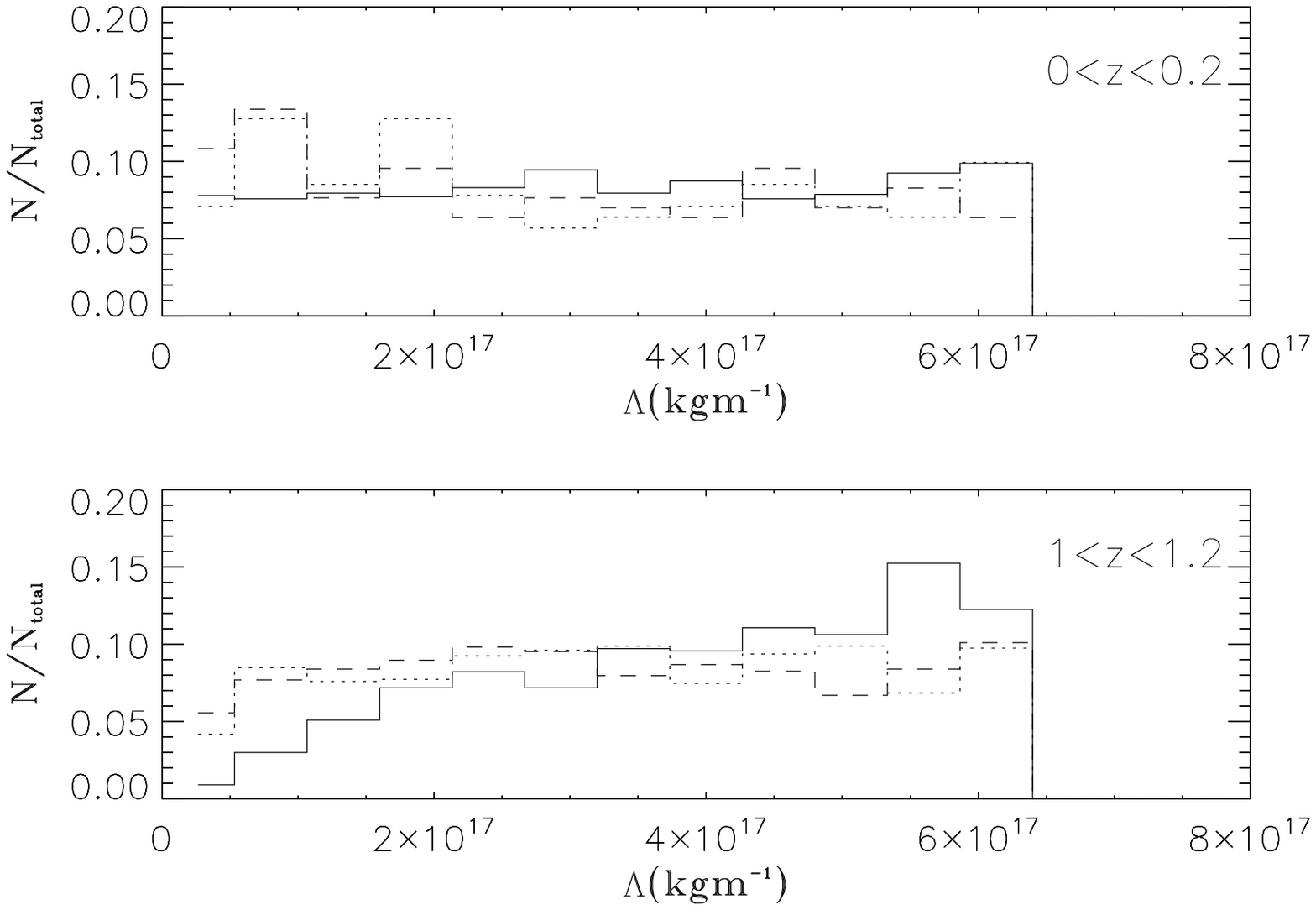}
\caption{Final $\Lambda$ distribution of the artificial samples generated by Model C in two redshift ranges. The upper plane is for $0\leq z\leq 0.2$ and the lower plane is for $1.0\leq z\leq
1.2$. $N_{\rm total}$ is the total source number in each redshift bin.
The solid line is for 3CRR, the dotted line for 6CE and the dashed line for 7CRS.}\label{n_lambda}
\end{figure}

In constructing our artificial samples we choose random values for the density parameter $\Lambda$ and the source age $t$ from uniform distributions. The jet power $Q_0$ is then adjusted to give the model source the correct radio luminosity. Not every possible combination of the set of three parameters $\Lambda$, $t$ and $Q_0$ is allowed because of restrictions on the magnitude of $Q_0$ and the selection criteria of the observed sample we compare with. It is
therefore not clear {\em a priori}\/ whether the distributions of $\Lambda$ and $t$ amongst the objects within the final artificial sample are also uniform. Any deviation from a uniform distribution in the final sample may reveal a genuine property of the source environments or source ages in the universe.

We first investigate the distribution of $\Lambda$ in the artificial sample arising from Model C. Figure \ref{n_lambda} shows the binned distribution of $\Lambda$ for two different redshift ranges, $0 \le z \le 0.2$ and $1 \le z \le 1.2$. At low redshifts the distribution is uniform within the fluctuations arising from the finite number of
sources in the artificial samples, regardless of which observed sample we compare with. At high redshifts the distribution of $\Lambda$ in the artificial sample mimicking the 3CRR sample is biased towards large values. The distributions within the artificial samples compared with 6CE and 7CRS remain uniform except for a drop in the bin of the smallest density. This behaviour is caused by the flux limits of the samples. Objects located in denser environments are more luminous. At low redshifts the flux limit of all samples corresponds to such low luminosities that the entire radio source population with an FRII-type morphology is represented in the samples. However, at higher redshifts the flux limit excludes sources in less dense environments as their luminosity is too low. Samples with a lower flux limit like 6CE and 7CRS obviously suffer less from this problem.

The result for the final distribution of $\Lambda$ arising from an initially uniform distribution implies that our current data is insufficient to constrain the distribution of the environmental properties of FRII-type sources. The uniform distribution `survives' unchanged through the source selection process. However, the distribution of $\Lambda$ in the universe itself may be uniform and thus, by chance, we have selected the correct distribution. To test this we also generate artificial samples with a Gaussian distribution and also a fixed value of $\Lambda$. As discussed in Section 8.3, the artificial samples using these two alternatives agree with the observed samples to a similar degree as the standard Model C with a uniform distribution. The final distribution of $\Lambda$ in the artificial sample drawing random values from a Gaussian distribution is also Gaussian. By construction, all the
sources in the final sample with a fixed value of $\Lambda$ are assigned this one value. From these results we conclude that the distribution of $\Lambda$ in the universe cannot be constrained with the currently available data.

\begin{figure}
\includegraphics[width=0.5\textwidth]{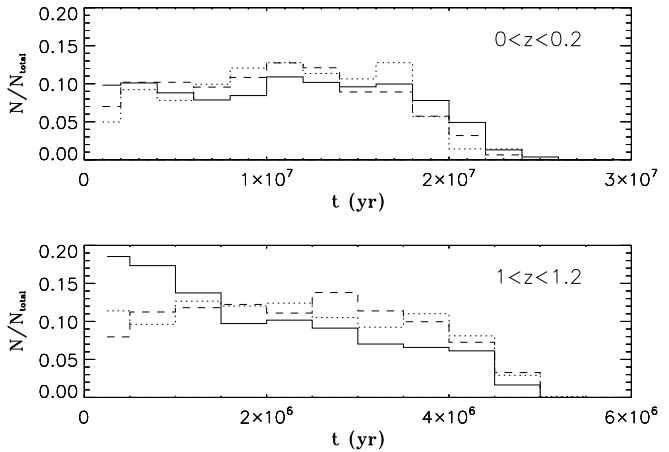}
\caption{Final distribution of $t$ in the artificial samples generated with Model C in two redshift ranges. The upper plane is
for $0\leq z\leq 0.2$ and the lower plane is for $1.0\leq z\leq
1.2$. $N_{\rm total}$ is the total source number in each redshift bin.
The solid line is for 3CRR, the dotted line for 6CE and the dashed line for 7CRS.}\label{n_t0}
\end{figure}

The distribution of source ages in our artificial samples for two redshift bins is shown in Figure \ref{n_t0}. The distributions at low redshift are uniform with a slow decrease for old ages. The decrease is caused by two effects. For Model C the source lifetime is decreasing with increasing redshift. Hence within the respective redshift ranges there is also by construction a range of source lifetimes. Sources towards the high redshift end of the range cannot
contribute to the bins of $t$ corresponding to the oldest sources. The second effect is again due to the flux limit of the samples. The radio luminosity of sources decreases as they get older. Therefore older sources are less likely to be included in the samples \citep[see also][]{b25,b14}.

At higher redshifts the high flux limit of the 3CRR sample leads to an age distribution skewed towards younger ages for which the sources are more luminous. This effect plays no significant role for the 6CE and 7CRS samples because of their lower flux limits. The drop in the source numbers in the bin containing the youngest objects is caused by the limit we impose on the size of sources to be included in the artificial samples.

The uniform distribution of $t$ in the artificial samples is consistent with our assumption of a maximum lifetime common to all sources at a given redshift. Any deviations from a uniform distribution, apart from those discussed above, would imply that our assumption is flawed. However, we cannot turn this argument around. The fact that the uniform distribution of $t$ `survives' the source selection in our model only demonstrates that, similarly to the
situation with the distribution of $\Lambda$, the current data does not constrain the distribution of source lifetimes. We cannot conclude that $t_{\rm{max}}$ is the same for all sources at a given redshift. 

\begin{figure}
\includegraphics[width=0.5\textwidth]{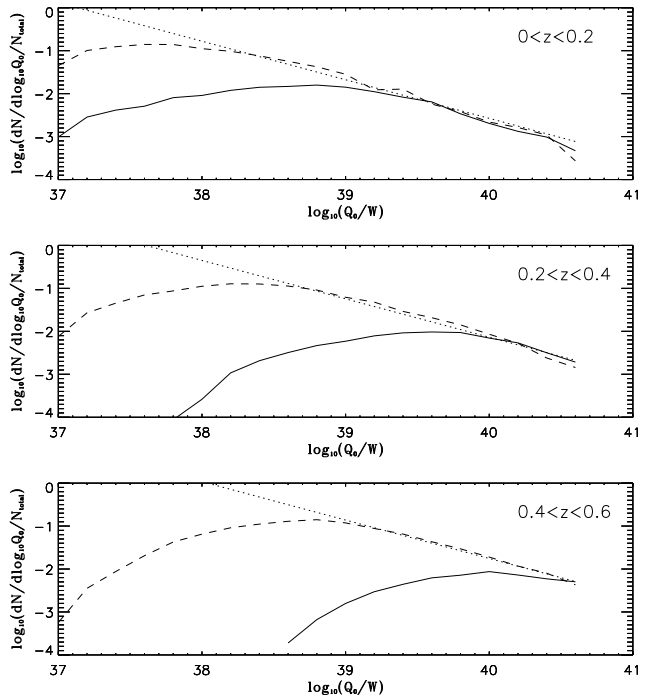}
\caption{The final distribution of $Q_{0}$ in three redshift ranges.
$N_{\rm total}$ is the total source number in each redshift bin. The
solid line is for 3CRR, the dashed line for 7CRS and the dotted line
shows a power-law with exponent $-1$. We have normalized the number
of the sources so that the curves at the high power end of the
artificial samples equivalent to 3CRR and 7CRS are
aligned.}\label{n_q}
\end{figure}

We do not initially constrain the distribution of the jet power, $Q_0$, to take a specific form as we do with $\Lambda$ and $t$. For each source we adjust $Q_0$ to give the correct radio luminosity using the randomly selected values for $\Lambda$ and $t$. The only restrictions we apply are that $Q_0$ cannot lie outside the range from $10^{37}$\,W to $5 \times 10^{40}$\,W. In this way the resulting final distribution for $Q_0$ arises from the constraints of the observed samples themselves. Figure \ref{n_q} shows the results for the artificial samples related to the 3CRR and 7CRS samples for three redshift ranges. We do not consider the result for the 6CE sample here as this sample also has an upper flux limit which affects the distribution of $Q_0$ at the high power end.

The shape of the distribution at the low power end is determined by the flux limit of the observed samples. Weaker jets produce less luminous lobes below the flux limit unless the sources are located in very dense environments and/or they are young. For high jet powers we expect that virtually all objects are above the sample flux limit, independent of environment or age, and that the $Q_0$ distributions shown in Figure \ref{n_q} are representative of the entire radio source population in the universe. This is supported by the agreement in the slope of the distributions for the 3CRR and the 7CRS samples in this region. The distribution for the artificial sample corresponding to the 3CRR sample turns over for higher jet powers because the flux limit for this sample is higher. We do not show results for redshifts beyond $z=0.6$, because the flux limit influences the high power end of the distribution, even for 7CRS.

The high power part of the $Q_0$ distribution is well approximated by a power-law. We find that ${\rm d} N / {\rm d} \log Q_0 \propto Q_0^{-0.90}$ for all three redshift ranges shown in Figure \ref{n_q}. We do not find any evidence for a change of this power-law slope as a function of redshift. We have also tested whether the slope changes when we change some parameters in the simulation proccess. If we adopt $\beta=1.5$, we find an exponent of $-1.05$, and the artificial samples from the GLF give an exponent of $-0.89$. Again, if we adopt a different initial distribution for $\Lambda$ as discussed in Section 8.3. For the Gaussian distribution the slope of the $Q_0$ distribution becomes slightly steeper, i.e. ${\rm d} N / {\rm d} \log Q_0 \propto Q_0^{-0.95}$. For the extreme assumption of a single value of $\Lambda$ for all sources we find a change of the power-law exponent to $-1.3$. We did not attempt a formal fit of a power-law to the high power end of the $Q_0$ distribution as it is not clear below which value of $Q_0$ we need to exclude data points from the fit to avoid the flux limit effects. Given the problems in determining the power-law slope, we conclude that ${\rm d} N / {\rm d} \log Q_0 \propto Q_0^{-1}$.  This result does not depend on the uncertain value of $\beta$ or the plausible values for the parameters describing the RLF, or the assumed distribution of $\Lambda$ as long as very extreme assumptions, e.g. a single value of $\Lambda$ for all sources, are avoided.

Various other authors have tried to constrain the distribution of $Q_0$ as well. Most of these studies present values for ${\rm d} N / {\rm d} Q_0$ rather than ${\rm d} N / {\rm d} \log Q_0$. We can easily convert our result by noting that ${\rm d} N / {\rm d} Q_0 = Q_0^{-1} {\rm d} N / {\rm d} \log Q_0 \propto Q_0^{-2}$. BRW from their work suggest that the power-law slope should be $-2.6$. More recently BW07 argued for a steeper exponent of $-3$ while KB07 find a value of $-1.6$. Our result is somewhat flatter than those of BRW and BW07 while it is slightly steeper than that of
KB07. The differences may be caused by the steeper luminosity evolution of the BRW model and the assumption of a single value for $\Lambda$ in BW07. KB07 ignored the effect of radiative energy losses of the synchrotron emitting
electrons on the luminosity evolution, which may explain the flatter distribution found by them.

The KDA model assumes that the energy contents of the magnetic field and of the relativistic, synchrotron radiation emitting particles in the lobe initially follow the minimum energy relation \citep[e.g.][]{b39}. Radiative energy losses change this situation somewhat for older sources, but the deviations of the model from minimum energy
conditions are small for most sources. Under minimum energy conditions the strength of the magnetic field and the volume of the emission region completely determine the radio luminosity. In our model the volume of the lobe is determined by its length and the energy density of the magnetic field is proportional to the lobe pressure, $p_{\rm c}$. Hence the measurements of lobe length and radio luminosity allow only a small range of possible lobe pressures. The data from the observed samples should therefore tightly constrain the distribution of lobe pressures in our
artificial samples.

\begin{figure}
\includegraphics[width=0.5\textwidth]{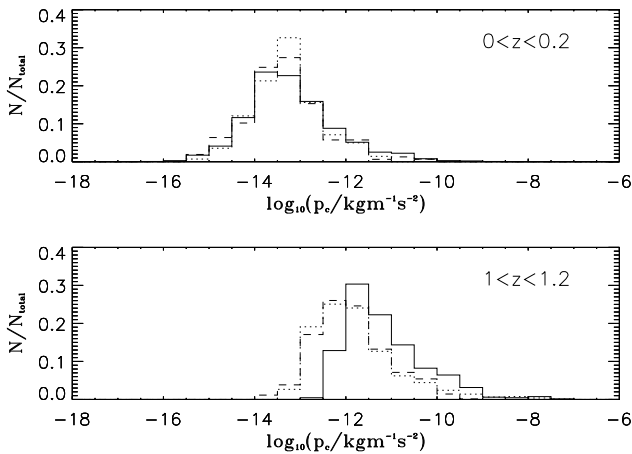}
\caption{Final distribution of lobe pressures, $p_{\rm c}$, in two redshift ranges. The upper plane is
for $0\leq z\leq 0.2$ and the lower plane is for $1.0\leq z\leq
1.2$. $N_{\rm total}$ is the total source number in each redshift bin. The
solid line is for 3CRR, the dotted line for 6CE and the dashed line for 7CRS.}\label{n_pc}
\end{figure}

Figure \ref{n_pc} shows the distribution of the lobe pressure for two different redshift ranges. The agreement between the three samples at low redshifts is good, indicating that the different flux limits do not influence the shape of the distribution. We take this as evidence that at low redshift we observe the entire FRII population. At high redshift the sample with the highest flux limit, 3CRR, shows a shift of the peak in the pressure distribution to
higher pressures compared to the other samples. Sources with lower lobe pressures are not luminous enough to be included in the sample and hence are missing. The agreement between the artificial samples corresponding to 6CE and 7CRS may indicate that these samples still include the entire source population at this higher redshift.

The pressure distribution is peaked and the peak shifts to higher values at higher redshifts. Our Model C implies that the maximum lifetime of a source decreases as $(1+z)^{-2.4}$ with increasing redshift. This also implies a proportional decrease of the average age of sources, $\left< t \right>$, included in the sample. Our implementation of the KA97 model predicts that $\left< p_{\rm c}\right> \propto \left<t\right>^{-2}$ and therefore $\left< p_{\rm
c} \right> \propto \left( 1 + z \right)^{4.8}$. We expect the average lobe pressure to increase by a factor of roughly 22 (1.35 in the logarithmic scale used in Figure \ref{n_pc}) between redshifts $z=0.1$ and $z=1.1$. This is consistent with the shift of the peak in the pressure distributions for the samples with the lower flux limit in the Figure \ref{n_pc}.

Given the good agreement of the pressure distributions between the artificial samples corresponding to 6CE and 7CRS in both redshift ranges, we argue that the samples cover the entire FRII source population at both redshifts without excluding too many sources through the respective flux limits. If so, then the shift of the peak in the pressure distribution with redshift is evidence for cosmological evolution of the population. This conclusion stems from
our assumption of conditions close to those described by minimum energy assumptions inside the lobes. It does not depend on the additional assumption of Model C of a decreasing maximum lifetime of sources with increasing redshift. The application of Model B would result in the same shift in the pressure peak. Unless we invoke a systematic change with redshift away from minimum energy conditions inside the lobes, the data imply that the average lobe pressure is
increasing rapidly with redshift out to about $z=1$. Beyond this redshift the flux limits of the samples used here exclude some of the FRII-type objects and so we cannot determine whether this trend continues. Also, the current data do not allow us to determine whether this increase in pressure with redshift is due to decreasing source lifetime and/or to an increase in the density of the surrounding medium.

BRW and BW06 use a similar approach to constraining the radio source population. We extend the work of BW06 by testing our model predictions in the complete $P$-$D$-$z$ data cube. We do not consider the spectral index as the evolutionary models of FRII sources themselves restrict the possible range of the spectral index. Our result is similar to that of BW06 in that the BRW model gives the worst agreement. However, between the other two models, BW06 prefer the MK model as it produces a better description of the source number ratios at different redshifts. In our work, we use the RLF to avoid the fitting of number ratios and the KDA model gives a slightly better fit than the MK model. The KDA model is also simpler than the MK model and so we prefer the KDA model in this paper. Using the RLF, we do not need to constrain the distribution of jet powers as BRW and BW06 were required to do. The final distribution of $Q_{0}$ in our artificial samples agrees with a power-law distribution with an exponent of approximately $-2$, which is flatter than the assumption of BRW and the best fit values found by BW06.

\section{Conclusions}

We have constructed a method for generating artificial flux-limited radio samples. We use these samples to study the cosmological evolution of the FRII source population by comparing with observed samples. We use three different models for the evolution of the linear size and  the radio emission of individual FRII sources from KDA, BRW and MK. Comparing artificial with observed samples, the 3-D KS test indicates that the KDA model gives the best prediction. The MK model can give acceptable predictions, but it is more complex with more model parameters and it does not
significantly improve the fitting results. The BRW model gives the worst prediction because the radio luminosity decreases too fast as the FRII source get older.

The properties of FRII sources are required to evolve with redshift in order for our artificial sample to fit the observations. For $\beta=2$, we introduce two models to meet the requirements:
\begin{enumerate}
\item[-] Model B: The maximum value of the environment density parameter $\Lambda$ evolves with redshift. We describe Model B with best fitting parameters by $t=r\times5\times10^{7}$\,yr, $\Lambda=r\times3.7\times10^{17}(1+z)^{5.8}$\,kg\,m$^{-1}$, where $r$ is a random number with uniform distribution between 0 and 1.\\
\item[-] Model C: The maximum jet age evolves with redshift. Similar to Model B, Model C can be expressed by $t=r\times2.7\times10^{7}(1+z)^{2.4}$\,yr, $\Lambda=r\times6.4\times10^{17}$\,kg\,m$^{-1}$.
\end{enumerate}
Both Model B and Model C produce artificial samples in agreement with the observed samples according to the 3-D KS test. We cannot decide whether a reduced maximum age or a denser environment is present at high redshift. Of course we cannot rule out a combination of these two effects. In fact, the very strong cosmological evolution of the density in the source environment required in Model B appears unrealistic. It is likely that a more moderate increase of $\Lambda$ combines with a reduction of the jet lifetimes to produce the observed effect.

Using our artificial samples, we study the distribution of the properties of FRII sources. The distributions of $\Lambda$ and $t$ are uniform, as we assume initially. The initial setting of the distribution 'survives' the source selection in our simulation and their real distribution cannot be constrained by the currently available data. Unlike $\Lambda$ and $t$, we do not initially constrain the distribution of $Q_{0}$, and a power law distribution arises naturally from the final artificial sample. We find a power law exponent of $x\approx-2$, and the slope shows no significant change at different redshifts up to $z=0.6$. We also study the distribution of the lobe pressure. The peak shifts to a higher value at higher redshift up to $z=1.2$. This shift arises from our assumption of conditions inside the lobes close to those expected from the minimum energy requirement. It does not depend on other details of our model.

\section*{Acknowledgments}

We thank Professor Steve Rawlings and Dr. Mark Lacy for clarifying some problems with the observational data. We also thank Dr. Catherine Brocksopp for help with the analysis of VLA data.



\bsp

\label{lastpage}

\end{document}